\documentclass[preprint,12pt]{elsarticle}

\usepackage[a4paper, total={6.7in, 10in}]{geometry}
\usepackage{amssymb}
\usepackage{amsthm}
\usepackage{amsmath}
\usepackage{epstopdf}
\usepackage{xcolor}
\usepackage{subcaption}
\usepackage{hyperref}
\usepackage{listings}

\journal{Chaos, Solitons and Fractals}

\begin{document}

\begin{frontmatter}

\title{The impact of social media on polarization in the society}
\author[label2]{Samana Pranesh}
\affiliation[label2]{organization={The Uncertainty Lab, Department of Applied Mechanics \& Biomedical Engineering, Indian Institute of Technology Madras},
           city={Chennai},
            postcode={600036}, 
            state={Tamil Nadu},
            country={India}}
\ead{psamana97@gmail.com}

\author[label2,label3]{Sayan Gupta\fnref{CA}}
\ead{sayan@iitm.ac.in}
\fntext[CA]{Corresponding Author}

\affiliation[label3]{organization={Complex Systems and Dynamics Group, Indian Institute of Technology Madras},
            city={Chennai},
            postcode={600036}, 
            state={Tamil Nadu},
            country={India}}

\begin{abstract}
The advent of social media platforms has revolutionized information consumption patterns, with individuals frequently engaging in these platforms for social interactions.
This trend has fostered an environment where people gravitate towards information that aligns with their preconceived notions, leading to the formation of echo chambers and polarization within the society. Recently introduced activity-driven models have been successful in capturing the dynamics of information propagation and polarization. The present study uses this model  to explore the impact of social media on a polarized society. By considering the varying influence of media, ranging from exposing individuals to contradictory views to reinforcing existing opinions, a supercritical pitchfork bifurcation is observed, triggering a transition from consensus to polarization. The transition points from polarization to consensus are derived analytically and  is validated through numerical simulations. This research sheds light on the complex interplay between social media dynamics and societal polarization.

\end{abstract}

\begin{highlights}
\item Investigates the effect of social media on a polarized society using activity-driven model.
\item Individuals are exposed to contradictory views as well as views which align with their existing belief. 
\item A supercritical pitchfork bifurcation is observed as the system transitions from consensus to polarization.
\item Transition points are derived analytically and validated through numerical simulations.

\end{highlights}

\begin{keyword}
Opinion dynamics \sep Polarization \sep Supercritical pitchfork bifurcation  \sep Social media.
\end{keyword}
\end{frontmatter}

\section{Introduction}
The increase in the usage of social media platforms like Facebook, Twitter and Instagram has not only changed the way people consume information \cite{moussaid2013social,lorenz2019accelerating} but also how information propagates through society, as this has given people the freedom to express themselves, communicate with other individuals around the world and participate in conversations. Social media platforms provide
individuals to voice their opinions, encourage inclusiveness and have democratized information flow, beyond official channels. Unfortunately, an unsavoury side to this free flow of information is that these platforms have not only enabled the spread of misinformation
\cite{hills2019dark,bessi2016homophily}  but coupled with  echo chambers~\cite{conover2011political,garrett2009echo,jamieson2008echo,sasahara2021social}, have enabled faster propagation of this misinformation. In social media parlance, echo chambers constitute a network of closely connected individuals who share common beliefs and serve to amplify information that is spread by one of their cronies. The existence of these cliques enable faster propagation of an idea or information through the society and can pose significant challenges to regulatory bodies, when the information have the potential to create societal unrest.  There is therefore significant interest in modelling the dynamics of information propagation through social media networks.
Few recent models of opinion dynamics~\cite{baumann2020modeling,baumann2021emergence,santos2021link} have been successful in capturing the features of echo chambers quantitatively. The 
recommendation engine of the online social media platforms accelerates the formation of echo chambers as it allows  users to interact with like minded peers~\cite{cinelli2021echo,del2016echo,cota2019quantifying,garimella2018political}.

Polarization and the formation of echo chambers in online social media platforms intensifies during debates involving controversial issues and significant political events~\cite{cardenal2019echo} such as 
elections~\cite{hanna2013partisan}, street protests~\cite{borge2015content}, and  societal issues 
that capture  public attention~\cite{barbera2015tweeting}. Although echo chambers are not permanent in social media, recent studies highlight the negative impact as they become the breeding ground for the spread of ``fake news"\cite{wang2020viral}. Several attempts have been made to prevent the formation of echo chambers. An introduction of random dynamical nudge depolarizes existing echo chambers and also help in preventing the formation of new echo chambers~\cite{currin2022depolarization}. This has been shown to be applicable for two well known models of opinion dynamics. Another recent study also showed that when individuals in an online social platform are slightly nudged to form new connections, it significantly avoids the formation of polarization~\cite{pal2023depolarization}.

Opinion dynamics model can be classified as discrete or continuous based on the nature of its  state variable. Voter model~\cite{redner2019reality} is the simplest discrete model in which opinion of each individual is described as a binary state variable and is appropriate in a bi-party election. As there are only two states, the rules of the model is described by the transition rates~\cite{gleeson2011high}, the probability by which opinion of an individual changes from one state to another. This model has  been applied to heterogeneous networks~\cite{masuda2010heterogeneous,sood2005voter} and time-varying networks~\cite{benczik2009opinion,suchecki2004conservation}. Voter model has also been studied in the presence of noise~\cite{granovsky1995noisy}. Several attempts have been made to extend the two state variable models to multi state discrete variable models to make voter-like model represent realistic scenarios~\cite{herrerias2019consensus}. Surprisingly, multi-state variable opinion dynamics models behave in a similar way as that of two state models. 

In models with continuous opinions, an individual is not restricted to have opinions on one or more topics within a set of possibilities. Instead, the state variable is defined on 
a finite interval 
i.e., $x_i(t) \in [0,1]$ where $x_i(t)$ is the opinion of the $i$-th individual \cite{peralta2022opinion}. One of the popular opinion dynamics model is the Deffuant model~\cite{deffuant2000mixing}, which assumes that   a pair of individuals interact only if their opinions are similar, i.e., $|x_i(t) - x_j(t)| < \epsilon$. This model is also known as bounded confidence model~\cite{rainer2002opinion} and depending on $\epsilon$, this model can reproduce consensus, bimodal or multi-modal opinion distributions. Bounded confidence model has  been extended to networks~\cite{fennell2021generalized}, multi-dimensional opinions~\cite{pedraza2021analytical} and its dynamics has been  studied in the presence of noise~\cite{carro2013role} and also neighbourhood effects \cite{krishnagopal2024bounded}. This model has also been extended to nonlinear case, called as nonlinear bounded confidence model and explored in the presence of both persuadable individuals and zealots \cite{brooks2024emergence}. Recently, studies focussed on opinion dynamics models for interdependent topics in the presence of individuals who are less prone to change their mind, referred to as zealots~\cite{ojer2023modeling}. This model has also been extended for the case of complex networks~\cite{ojer2023vanishing}. The opinion dynamics model introduced by Baumann et al.\cite{baumann2020modeling}  is an activity-driven model, and has been successful in reproducing the characteristics shown by empirical data. This model incorporates homophily factor, which is the tendency of the individuals to interact with people of similar opinions, and is able to capture consensus, radicalization and polarization. This activity-driven model has also been extended to bi-layer complex networks \cite{gajewski2022transitions}. A recent study has further refined this activity-driven model by assuming that the agents only interact with their neighbors directly, instead of distant or global agents \cite{dong2024evolution}. 

\begin{figure}[htbp]
\centering
    \includegraphics[width=5.7in,height=2.3in]{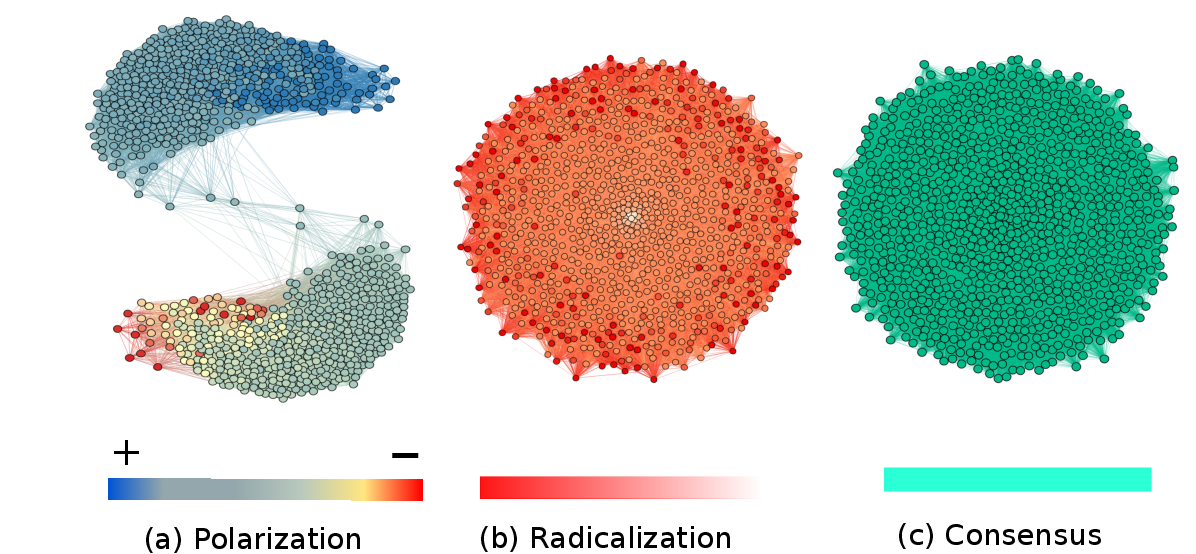}
    \caption{A schematic representation of (a) Polarization (b) Radicalization (c) Consensus; in a network of $1000$ individuals. Color of the nodes in the network represent the opinion of each individual on a given issue.  }
    \label{fig:gephi}
\end{figure}

A graphical representation of states of polarization, radicalization and consensus in a network of $1000$ individuals is shown in Fig. \ref{fig:gephi}(a), (b) and (c) respectively. 
Here, the state is taken to be continuous.
A network that is polarized consists of
two distinct clusters representing the two opposite opinions on a given issue. Extreme opinions on either side is represented by darker shades (of blue and red),  with the extremities of the state  marked as $+$ and $-$ in the colorbar of  Fig. \ref{fig:gephi}(a). Within each clusters, despite a collective inclination towards a specific viewpoint, the intensity of an individual  opinion varies. This spectrum of intensity is visualized through the gradation of color, with darker tones of blue or red signifying  extreme opinions within each cluster. In the case of radicalization, the opinions of the individuals are absorbed by either side of the viewpoint, shown in varying shades of red in Fig. \ref{fig:gephi}(b) whereas in consensus, the individuals remain neutral and do not have an opinion on a given issue, shown schematically in green in Fig.\ref{fig:gephi}(c). 

The present study utilizes the mathematical model introduced by Baumann et al.\cite{baumann2020modeling} to examine how  selective exposure of an individual to media content impacts opinion formation. Selective exposure, where individuals favor media aligning with their emotional and cognitive preferences, often leads to confirmation bias \cite{nickerson1998confirmation}, a key driver of opinion polarization. Furthermore, this study delves into scenarios where individuals actively seek information contradicting their own perspectives. Through analytical and numerical studies, it is shown that society undergoes a transition from polarized to a consensus state when exposed to opposing views. The paper is structured as follows: Section \ref{mathematical_model} provides a background on the activity-driven model which is used in this study.  Results and related discussions are presented in Section \ref{results}. The salient outcomes of this study are summarized in the concluding section in Section \ref{conclusion}.

\section{Mathematical model} \label{mathematical_model}
The activity-driven (AD) model,  proposed by Baumann et al. \cite{baumann2020modeling} is used for exploring the influence of social media on polarization within society. This model effectively captures certain aspects of opinion dynamics, such as the formation of echo chambers and polarization as a result of interactions in a network of individuals. 

\subsection{Background}
The governing equations for the evolution of opinions according to the AD model, is given by
\begin{align} \label{equation1}
\dot{x}_i = -x_i +  \left[K \sum_{j=1}^{N} A_{ij}(t) \tanh(\alpha x_j)\right] .
\end{align}
Here, each of the $N$ agents in a network, denoted as $i$, represent a time evolving opinion (state variable) denoted by $x_i(t) \in [-\infty, \infty]$, $K > 0$ is the interaction strength between any two agents $i$ and $j$, and $\alpha > 0$ denotes \textit{controversialness} \cite{baumann2020modeling} of a given topic. Additionally, this model assumes that each topic can attain polarization 
into two opposing stands - for or against. The stance taken by each agent is determined by the sign of $x_i$, and is mathematically denoted by the signum function 
$\sigma(x_i)$. The magnitude of $x_i$ reflects the strength of conviction of an agent towards the topic.
The interactions among the agents are described through the temporal adjacency matrix denoted as $A_{ij}(t)$. Specifically, $A_{ij}(t) = 1$ signifies an input from agent $j$ to agent $i$, while $A_{ij}(t) = 0$ indicates no such input. The probabilistic factor of reciprocity, $r \in [0, 1]$, determines the likelihood that a connection between agents is mutually influential. When $A_{ij}(t) = A_{ji}(t) = 1$ due to reciprocity, both agents participate in updating their opinions based on this interaction. Conversely, if the interaction lacks reciprocity (i.e., only one direction $A_{ij}(t) = 1$ or $A_{ji}(t) = 1$), the connection is unidirectional and an update of opinions occurs  only in the agent downstream of the connection. 
Additionally, in the $AD$ model~\cite{perra2012activity,starnini2013topological}, each agent possess an activity level, denoted by $a_i$ and only active agents can interact with other individuals. This factor is assumed to be different for each individual and therefore follows a distribution. 
Empirical data reveals that the probability distribution of $a_i$ follows a power-law  \cite{perra2012activity, moinet2015burstiness},  given by $F(a) = \frac{1 - \gamma}{1-\varepsilon^{1-\gamma}} a^{-\gamma}$. The parameter $\gamma$ is the power-law exponent which controls how steeply the activity probability distribution decreases. If $a_i \in [\epsilon, 1]$ at time $t$, where $\varepsilon$ denotes the minimum activity level, the agent interacts with $m$ other individuals based on a weighted probability $p_{ij}$ given by, $p_{ij} = \frac{|x_i - x_j|^{-\beta}}{\sum_j |x_i - x_j|^{-\beta}}$. Here, the parameter $\beta$ serves as the homophily factor, signifying the inclination of agents with similar opinions to interact with each other. Specifically, when $\beta = 0$, there is no preference for interaction based on similarity of opinions. However, for $\beta > 0$, it implies that agents sharing similar viewpoints are more inclined to interact. The weighted probability is formulated to depict a power-law decay of connection probabilities. This formulation results in a situation where there is only a minimal probability for agents with divergent opinions to engage in interactions. 
Thus, $(\varepsilon, \beta, \gamma, m)$ capture the activity-driven model of opinion dynamics. 
\begin{figure}[htbp]
\centering
    \includegraphics[width=0.22\textwidth]{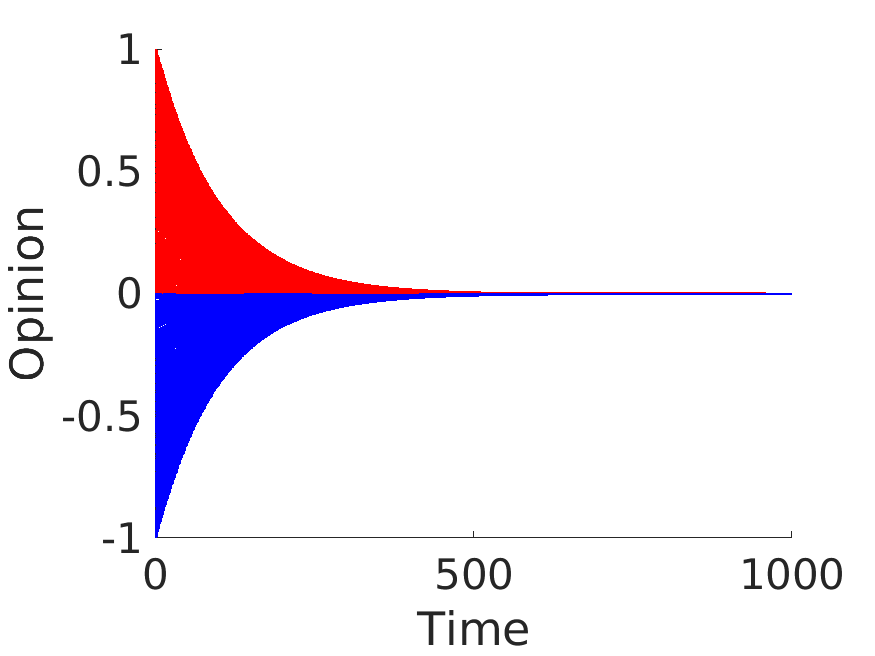}
    \includegraphics[width=0.22\textwidth]{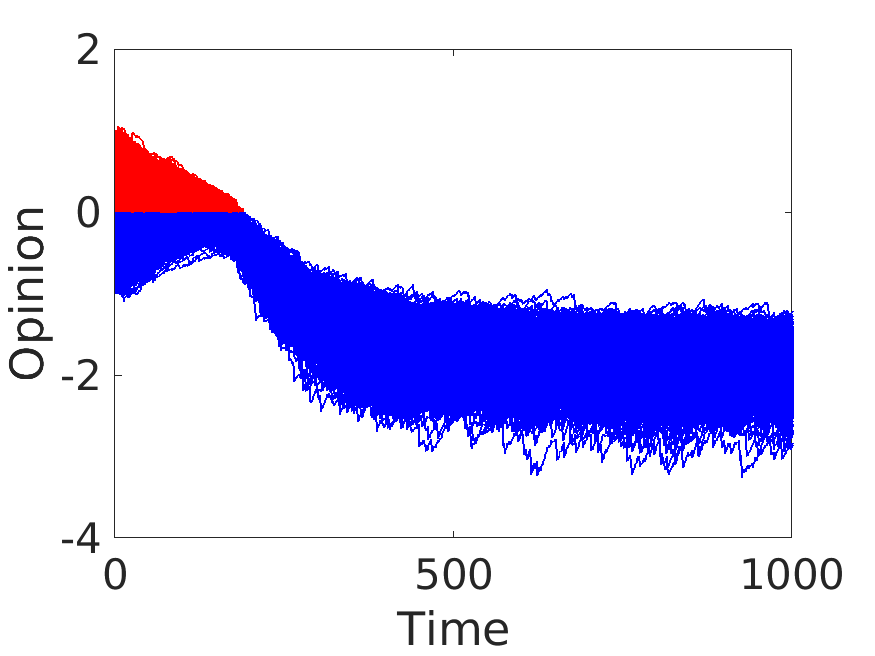}
    \includegraphics[width=0.22\textwidth]{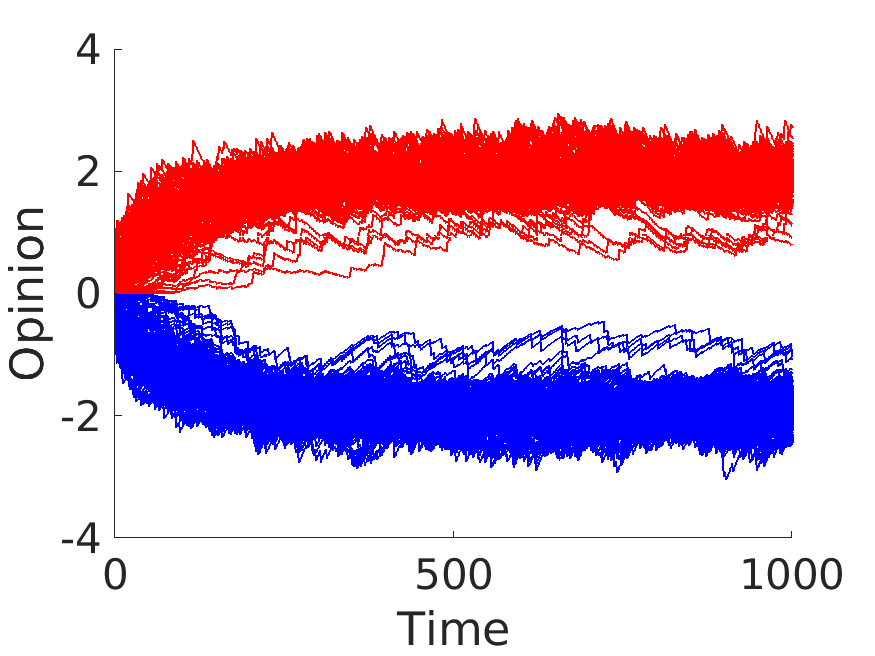}\\
    \textbf{(a)} \hspace{3.5cm} \textbf{(b)}\hspace{3.5cm} \textbf{(c)}\\
    \includegraphics[width=0.22\textwidth]{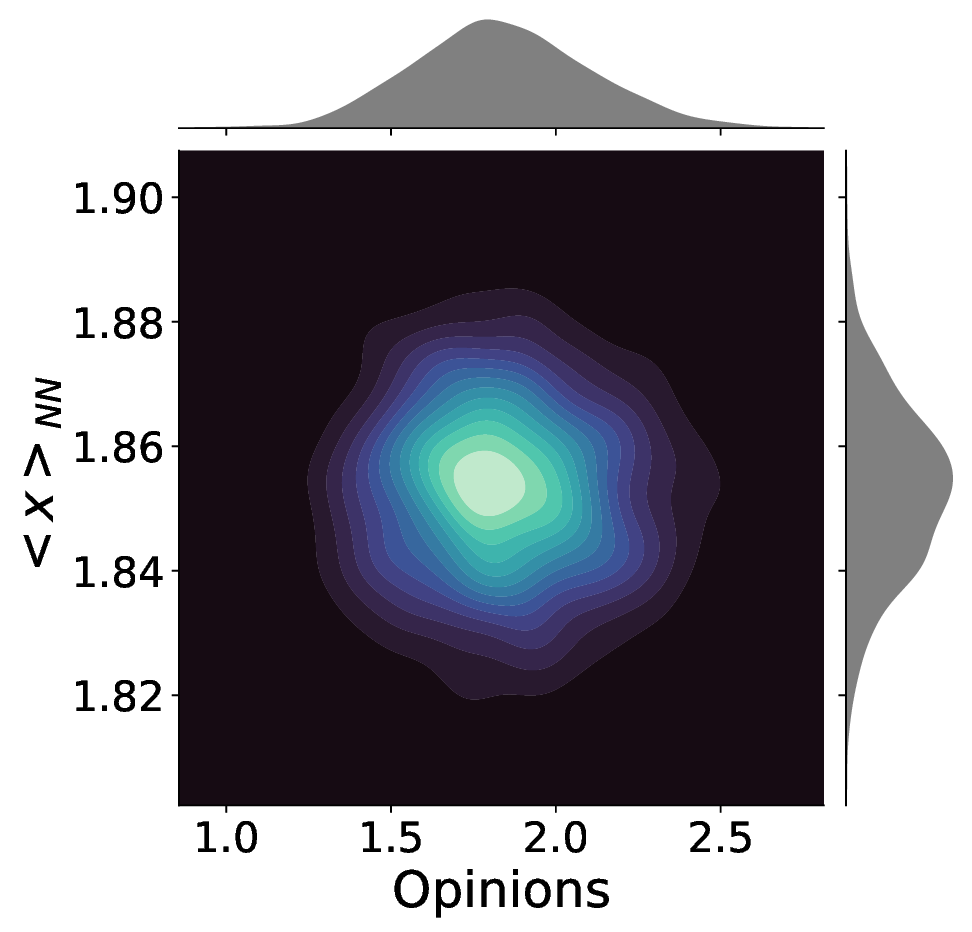}
    \includegraphics[width=0.22\textwidth]{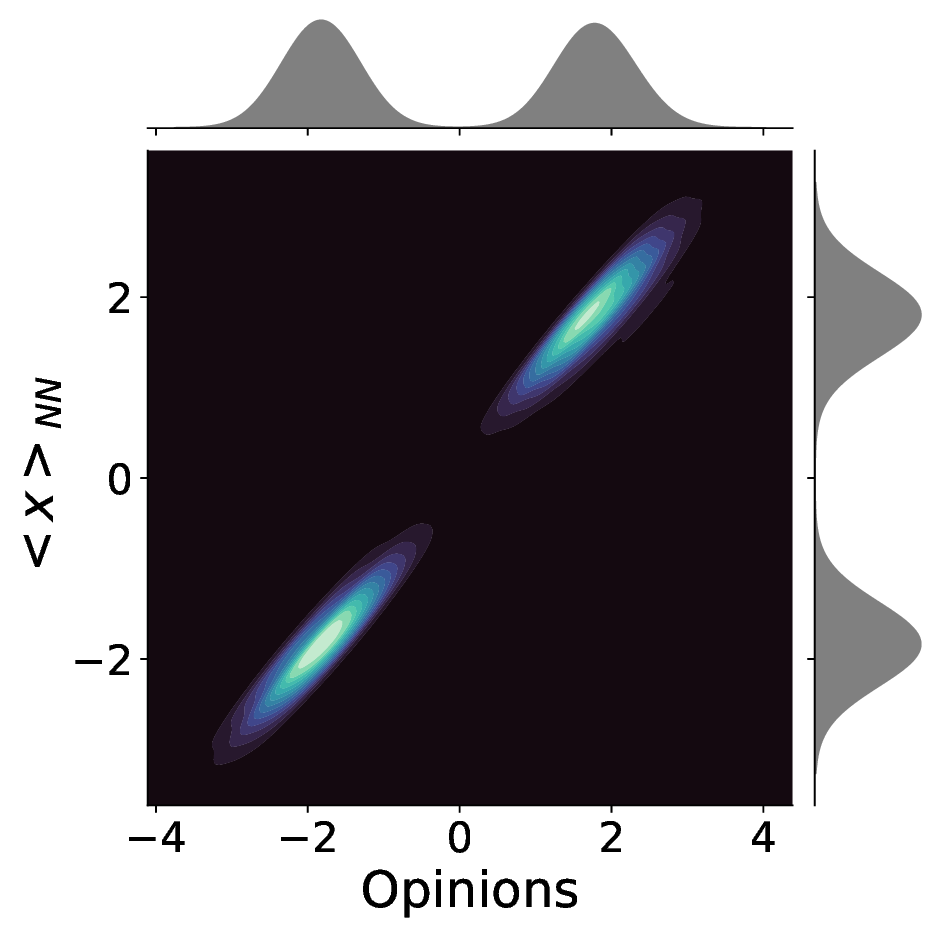}\\
    \textbf{(d)} \hspace{3.5cm} \textbf{(e)}
    \caption{Temporal evolution of the agents’ opinions. (a) Neutral consensus for which all opinions converge to zero
($\alpha = 0.05, \beta = 2, F_m = 0$). (b) (One-sided) radicalization $(\alpha = 3, \beta = 0, F_m = 0)$. (c) Opinion polarization, in which opinions split into two opposite sides $(\alpha = 3, \beta = 3, F_m = 0)$. Social interaction strength and reciprocity were set to $K = 3$ and $r = 0.5$, respectively (d-e) Heat maps of
steady state opinions versus mean of agents’ nearest neighbours opinions $\langle x \rangle_{NN}$ ; A unimodal distribution indicates radicalization and all the agents have similar opinion (see (d)) A bimodal distribution indicate the formation of echo chambers.
Agents interact with other agents who have similar opinions (see (e). The marginal distribution of opinions, $P(x)$, and average opinions of the nearest-neighbor $P(\langle x \rangle_{NN})$ are plotted on the x- and y-axis of (d) and (e), respectively. }
    \label{fig:fig1}
\end{figure}\\
When active agents have an equal probability of interacting with $m$ other agents, the network has the potential to radicalize, resulting in all agents adopting the same stance (as illustrated in Fig. \ref{fig:fig1}(b)). Conversely, if agents with similar opinions interact, this bias favours polarization of viewpoints and leads to the creation of echo chambers, as shown in Fig. \ref{fig:fig1}(c). In the absence of any interactions ($K=0$), the individuals maintain a neutral opinion; see Fig. \ref{fig:fig1}(a). When users tend to establish connections more frequently with peers who hold similar opinions, it leads to the formation of echo chambers, facilitating the exchange of information among like-minded individuals. This phenomenon, observed at a network scale, is characterized by a correlation between an individual's opinion $x_i$ and the average opinions held by their closest neighbors, and is mathematically given by $\langle x_i \rangle_{NN} = \frac{1}{k_i}\sum _j a_{ij}x_j$ \cite{baumann2020modeling}. Here, $a_{ij}$ represents the elements of the adjacency matrix obtained after Eq. \ref{equation1} is integrated for a long time so that steady state condition is achieved and $k_i$ is the total number of connections of agent $i$. Figs.\ref{fig:fig1}(d)-(e) show colored contour maps of the density of users in $x-\langle x \rangle_{NN}$ plane. The resulting bimodal distribution as shown in Fig.\ref{fig:fig1}(e) indicates the formation of echo chambers, 
 whereas a unimodal distribution shown in Fig.\ref{fig:fig1}(d) indicates that the agents share a similar opinion and depicts radicalization in the society. Note that radicalization is a distinct state from consensus, with the latter state being characterized by no significant difference in the state variables within the network. 

\subsection{Social media effects}
To take into account the effect of social media on a group of individuals who have tendency to be polarized, Eq.\eqref{equation1} is modified as 
\begin{align} \label{media}
\dot{x}_i = -x_i +  \left[K \sum_{j=1}^{N} A_{ij}(t) \tanh(\alpha x_j)\right] + F_m \tanh(\alpha_m x_i). 
\end{align}
The third term in Eq.\eqref{media} describes the influence of selective information from the media, which is a central focus of this study. This term in the equation is inspired by everyday experiences on social media platforms, such as video recommendations on YouTube. When one watches a video, the platform tends to suggest similar ones based on one's viewing history and this can have adverse effects like intensifying political polarization \cite{cho2020search}. This is modeled as $F_m \cdot f(x_i)$, where $F_m$ is a constant, denoting the strength or the magnitude of impact that the social media exerts on the opinions of an agent and signifies various factors such as the persuasive power, prominence, or intensity of the social media's influence on the opinion of the agents.  $f(x_i)$ is assumed to be a tan hyperbolic function, commonly chosen due to its symmetric nature \cite{baumann2020modeling, lee2022effect, gajewski2022transitions}.  
The parameter $\alpha_m$ is the willingness (converse of resistance) of individuals to be influenced by social media and reflects the degree to which the influence of the social media affects the viewpoint of the individuals. 

When $F_m>0$, the second term in Eq.\eqref{media} reinforces an agent's own opinion. Conversely, for $F_m<0$, this term operates inversely, leading to a scenario where positive (negative) values of $x_i$ result in a decrease (increase) in $x_i$. This indicates that media tends to either strengthen or weaken the opinion of each agent, depending on the sign of $F_m$.
As discussed in \cite{baumann2020modeling}, $A_{ij}(t)$ in Eq.\eqref{media} is approximated by its time average $\langle A_{ij}(t) \rangle_t = \lim_{T \to \infty} \frac{1}{T} \int_{0}^{T} A_{ij}(t) {\rm d}t$. By taking the time average, it is assumed that the rapid dynamics and fluctuations within the network average out or become negligible.  Eq.\eqref{media} can  now be rewritten as 
\begin{align} \label{modified_eqn}
\dot{x}_i = -x_i +  \left[K \sum_{j=1}^{N} \langle A_{ij}(t) \rangle_t \tanh(\alpha x_j)\right] + F_m \tanh(\alpha_m x_i) .
\end{align}
The $ij$-th element of the time-averaged adjacency matrix is,
$\langle A_{ij}(t) \rangle_t =\frac{m}{N}(ra_i + a_j)$\cite{baumann2020modeling},
which when averaged over all the activities in the system can be written as
\begin{align} \label{equation4}
c =\frac{m}{N}(1+r)\langle a \rangle .
\end{align}
Here, $\langle a \rangle$ is the average activities which is calculated by taking the first moment of $F(a)$:
\begin{align}
\langle a \rangle = \int_{1}^{\varepsilon} a \cdot F(a) \, {\rm d}a.
\end{align}
The expression for $F(a)$ is substituted in the above equation and simplified to get $\langle a \rangle$ as, 
\begin{align}
    \langle a \rangle = \frac{1-\gamma}{2-\gamma} \frac{1 - \epsilon^{(2 - \gamma)}}{1 - \epsilon^{(1 - \gamma)}}.
\end{align}
Substituting Eq.\eqref{equation4} in Eq.\eqref{modified_eqn}, leads to
\begin{align} \label{final}
\dot{x}_i = -x_i +  \left[K c \sum_{j=1}^{N}  \tanh(\alpha x_j)\right] + F_m \tanh(\alpha_m x_i) .
\end{align}
 Eq.\eqref{media} is integrated numerically using fourth order Runge-Kutta method with a step size of 0.01 and initial opinions $x_i$ uniformly distributed in $[-1,1]$. The $AD$ parameters are set to $\gamma = 2.1$, $\epsilon = 10^{-2}$, $m=10$ and $r = 0.5$ unless mentioned otherwise. In the absence of effects of social media, i.e. when $F_m=0$, Eq.\eqref{media}  reproduces the results obtained in \cite{baumann2020modeling}. 

\section{Results} \label{results}
\begin{figure}[htbp]
\centering
\includegraphics[width=0.8\textwidth]{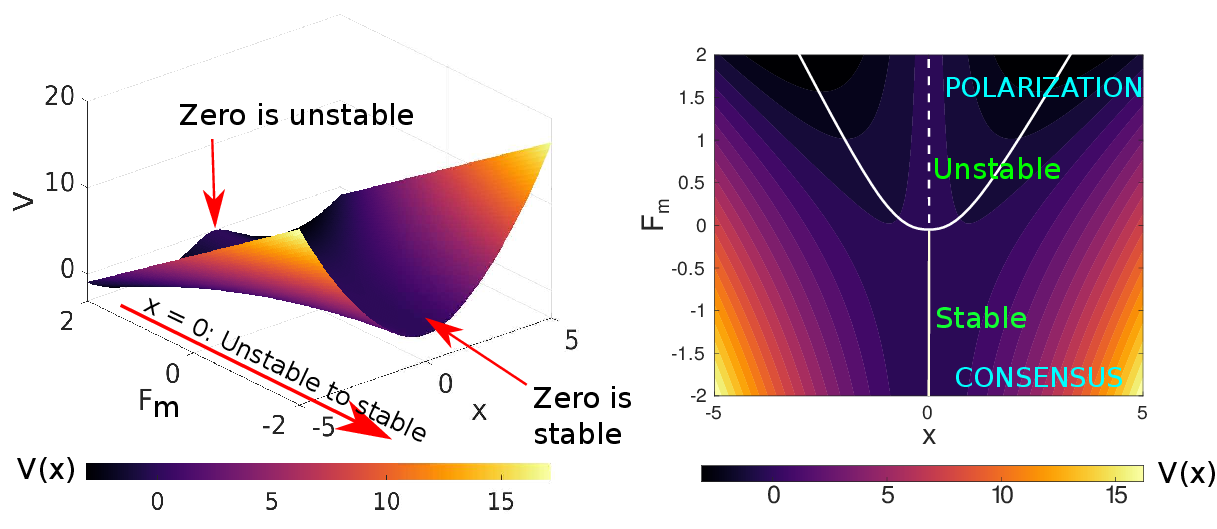}\\
\textbf{(a)} \hspace{7cm} \textbf{(b) }
\includegraphics[width=0.9\textwidth]{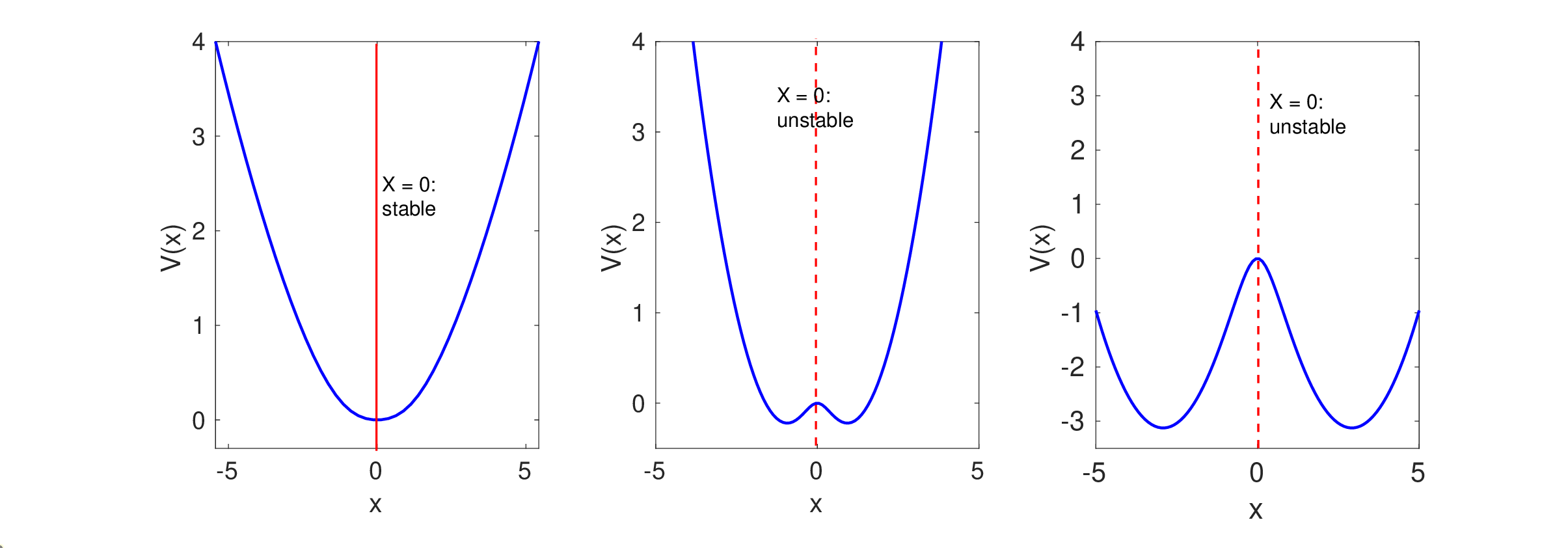}\\
\textbf{(c) $F_m = -2$} \hspace{2cm} \textbf{(d) $F_m = 0$} \hspace{2cm} \textbf{(e) $F_m = 2$}
    \caption{(a) Plot of potential $V$ as a function of $F_m$ and $x$. For positive values of $F_m$, $x=0$ is unstable whereas for negative values of $F_m$, $x=0$ is stable (b) Contour plot of potential: $x=0$ is stable for $F_m<0$ and unstable for $F_m>0$. Also, when $F_m>0$, there are two stable states. A stable state of zero indicates consensus, depicted by a solid white line, while the presence of two stable states is represented by solid white curved lines, accompanied by an unstable state at zero indicated by a dashed white line, defines the region of polarization (c)-(d)-(e)Plot of $V$ as a function of $x$ for $F_m=-2$, $F_m=0$ and $F_m=2$ respectively. As $F_m$ changes from negative to positive value, $x=0$ changes from stable to unstable along with the emergence of two stable states. The parameter values are: $\alpha=3$, $\beta=3$, $K=3$, $\alpha_m=0.5$. }
    \label{fig:potential}
\end{figure}

The  behaviour of the system described by Eq.\eqref{final} can be better understood by defining a  metric $V(x)$ and referred to as the potential function. For a first-order system of the form $\dot x = f(x)$, the potential $V(x)$ is defined as a scalar function, given by $f(x) = -\frac{{\rm d}V}{{\rm d}x}$ \cite{strogatz2018nonlinear} and can be viewed as an energy function, with its topology helping in describing the dynamical behavior of the system. The definition can be extended to higher order systems and the scalar definition of the function ensures a mapping from a multi-dimensional state space to the scalar space, enabling an easier appreciation of the overall dynamics of the system. Specifically, any point on the function where the gradient is zero defines an equilibrium point, whose stability can be examined by the nature of the gradient of the function in their neighborhood. Following this definition, the potential function for the system in Eq.\eqref{final} can be defined as
%
\cite{gajewski2022transitions}
\begin{align}
    V(x) = - \int_{-\infty}^{x} F(u) \, {\rm d}u,
\end{align}
where $F(u)$ is the effective force, appearing on  
the right-hand side of Eq.\eqref{final} and can be shown to be 
\begin{align}
    V(x) = \frac{x^2}{2} - \frac{K \cdot c}{\alpha} \log(\cosh{\alpha x}) - \frac{F_m}{\alpha_m} \log(\cosh{\alpha_m x}).
\end{align}
The plot in Fig. \ref{fig:potential}(a) illustrates the relationship between $V$, $F_m$, and $x$. An inspection of this figure reveals that the extrema - locations where the gradient of the potential function is zero - lies along the $x=0$ line. For regions when $F_m < 0$, the $x=0$ point lies in a valley and signifies stable equilibrium, while for regions when $F_m >0$, the point lies at the peak and corresponds to unstable configuration. In the latter case, while $x=0$ is unstable, there exists two additional stable equilibrium points. It is to be emphasized here that a  stable equilibrium at $x=0$ implies no further information propagation and hence corresponds to a stable state of consensus. 
 
 These features are better explained through the contour plot in the $F_m -x$ plane shown in 
 Fig. \ref{fig:potential}(b). Here, the full white lines represent the stable equilibrium points, while the dashed lines represent the unstable equilibria.
 This figure demonstrates that as $F_m$ transitions from negative to positive values, the stable consensus state at zero (marked in solid white line) becomes unstable, and simultaneously results in the birth of two additional stable equilibrium points on either side of $x=0$. The formation of these coexisting stable points implies polarization.
 This can also be observed from the potential plots  in Fig.\ref{fig:potential}(c), (d) and (e) for $F_m=-2$, $F_m=0$ and $F_m=2$ respectively. This transition mirrors the characteristics of a supercritical pitchfork bifurcation. Hence it can be concluded that, exposure to information conflicting with individual opinions tends to facilitate consensus formation, whereas gradual exposure to information aligning with preexisting beliefs triggers a supercritical pitchfork bifurcation, ultimately leading to polarization. In the subsequent sections, the cases where $F_m < 0$ and $F_m > 0$ are explored in detail to study the influence of social media on polarization prevailing in the society. 

\subsection{Negative media influence: $F_m < 0$}
\begin{figure}[htbp]
\centering
    \includegraphics[width=0.26\textwidth]{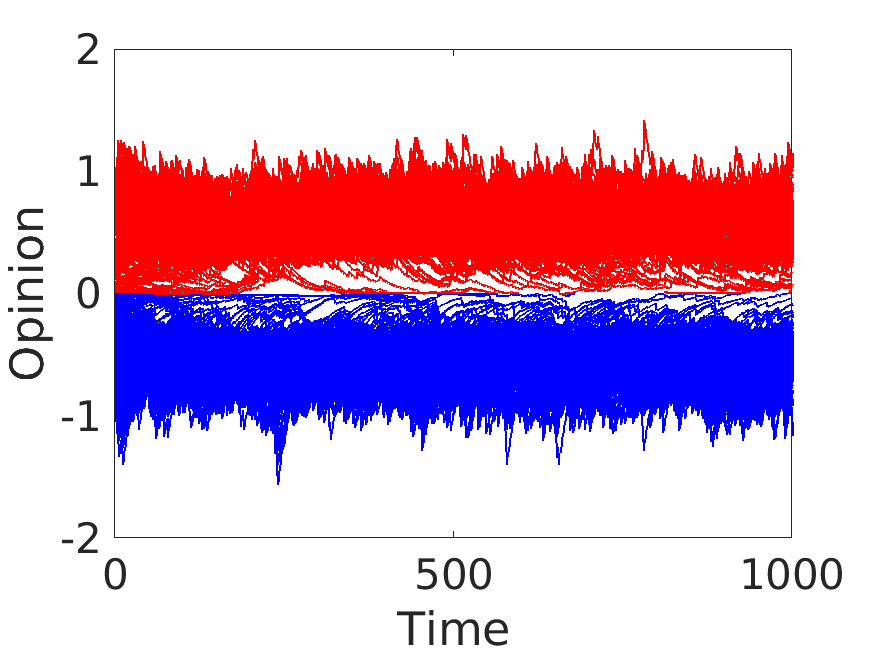}
    \includegraphics[width=0.26\textwidth]{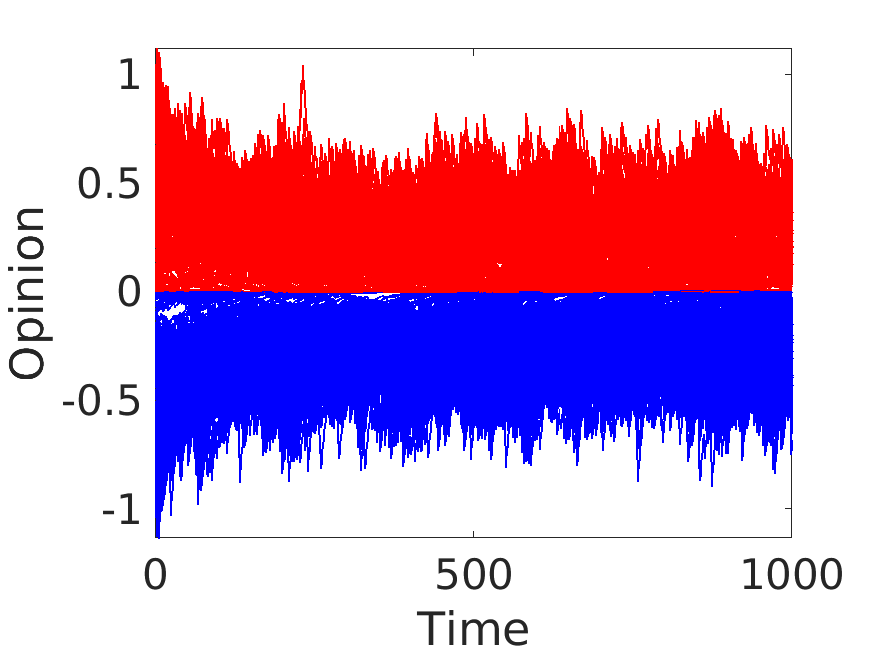}
    \includegraphics[width=0.26\textwidth]{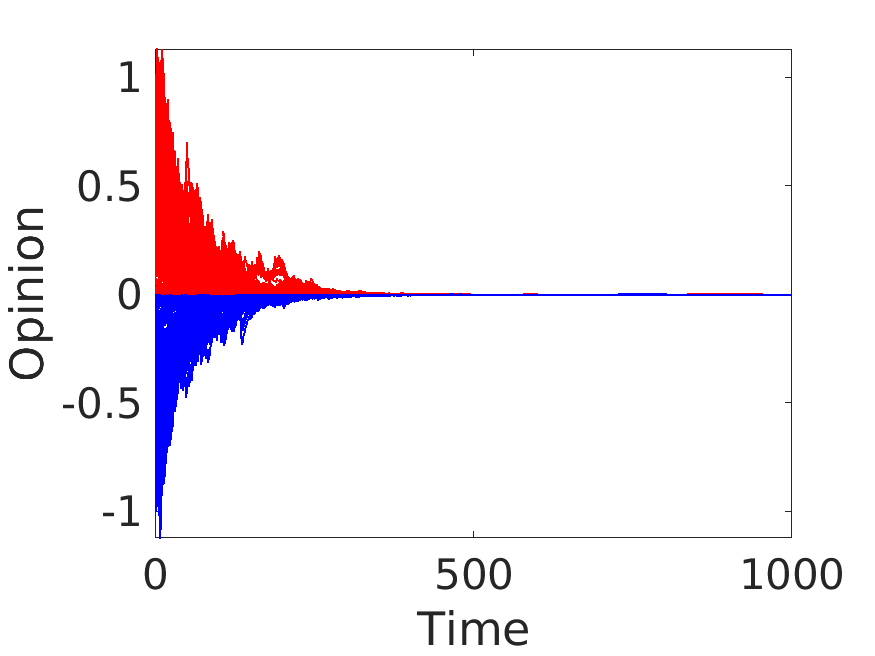}\\
    \textbf{(a)} \hspace{4.2cm} \textbf{(b)}\hspace{4.2cm} \textbf{(c)}\\
    \includegraphics[width=0.26\textwidth]{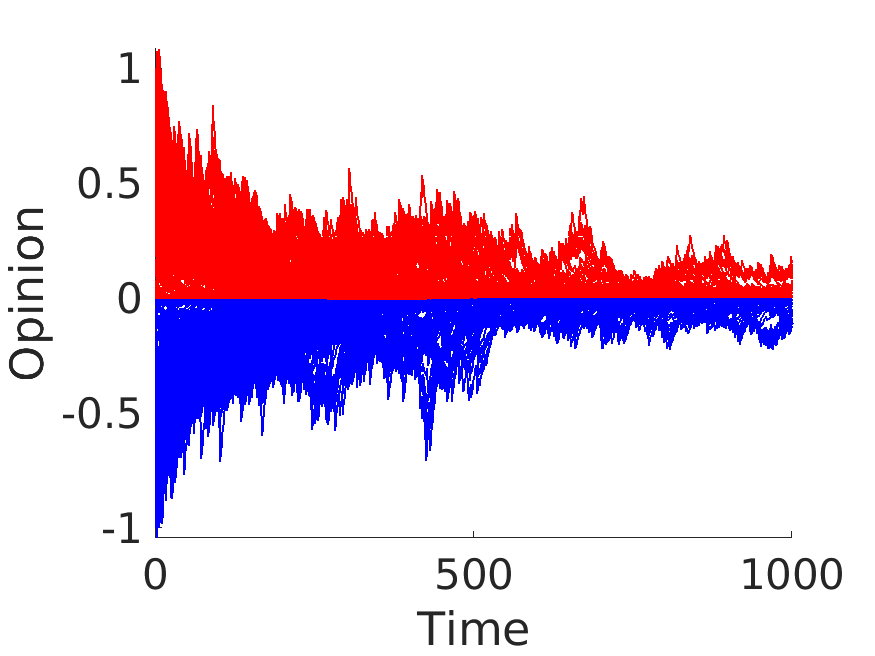}
    \includegraphics[width=0.26\textwidth]{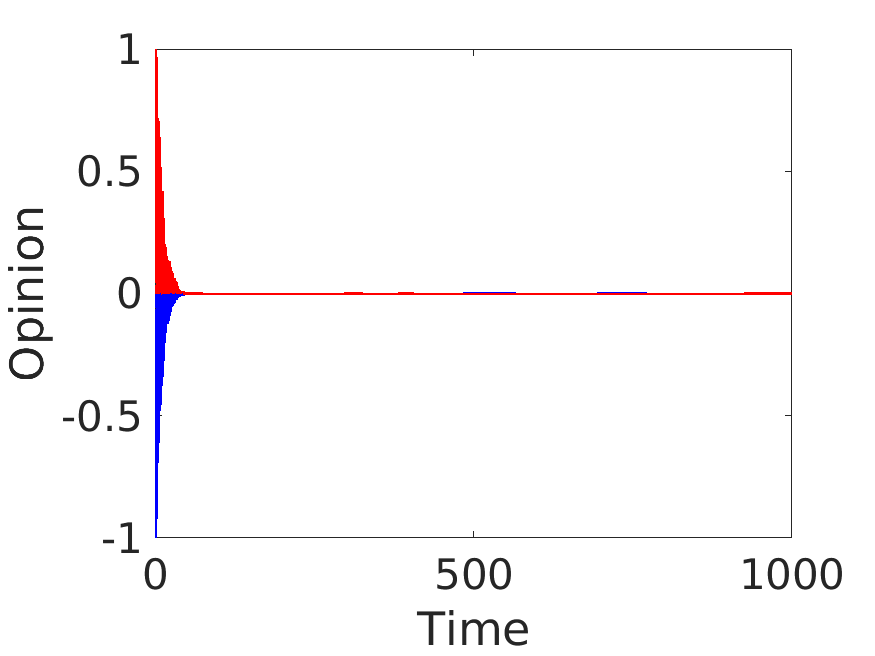}\\
    \textbf{(d)} \hspace{4.2cm} \textbf{(e)}\\
    \caption{Temporal evolution of agents' opinions for (a) $F_m = -0.3$, $\alpha_m = 0.07$ (b) $F_m = -0.5$, $\alpha_m = 0.07$ (c) $F_m = -0.9$, $\alpha_m = 0.07$ (d) $F_m = -0.9$, $\alpha_m = 0.05$ (e) $F_m = -0.9$, $\alpha_m = 0.2$. Other parameters are $\alpha = 3$ and $\beta = 3$. }
    \label{fig:fig2}
\end{figure}

This section analyzes the effects of negative social media influence (i.e. when $F_m < 0$) on a society which has the tendency to polarize. 
The numerical values of $\alpha$ and $\beta$ are fixed at $3$, which ensures that the society is in polarized state in the absence of external effects. Eq. \eqref{media} is numerically integrated for different values of $F_m$ and $\alpha_m$. Fig. \ref{fig:fig2} shows the temporal evolution of the opinion of all the agents, denoted by their state variable $x_i$, 
for different values of $\alpha_m$ and $F_m$. The top panel in Fig. \ref{fig:fig2} depicts the time histories {of $x_i$} for fixed values of $\alpha_m$ but for three distinct values of  $F_m$, while the bottom panel shows the temporal evolution {of the same} fixed values of $F_m=-0.9$ but two distinct values of $\alpha_m$.
From the top panel of Fig. \ref{fig:fig2}, it is observed that as $F_m$ becomes more negative for a fixed value of $\alpha_m$, the echo chambers depicted as two groups of time series - those with $x_i >0$ (shown in red) and those with $x_i<0$ (shown in blue), 
come closer to each other and beyond a critical value of $F_m$, the polarization  disappears with both clusters attaining the state of $x=0$, representing consensus; see Fig. \ref{fig:fig2}(c). 
From Figure \ref{fig:fig2} (d) and (e), it is seen that as the value of $\alpha_m$ is increased for a fixed value of $F_m$, the society reaches consensus at a much faster rate. 

A physical interpretation of these results are as follows:
A high negative media influence ($F_m << 0$) creates an environment where agents are exposed to  dominant but opposite narrative, and as a result become aware of diverse perspectives and gradually move towards a consensus. Additionally, a society where individuals are more open-minded (high value of $\alpha_m$) and willing to consider opposing views tends to reach a consensus more rapidly. This underscores the importance of individual openness and willingness to embrace diverse opinions in fostering societal unity and shared understanding.

\begin{figure}[t]
\centering
    \includegraphics[width=0.32\textwidth]{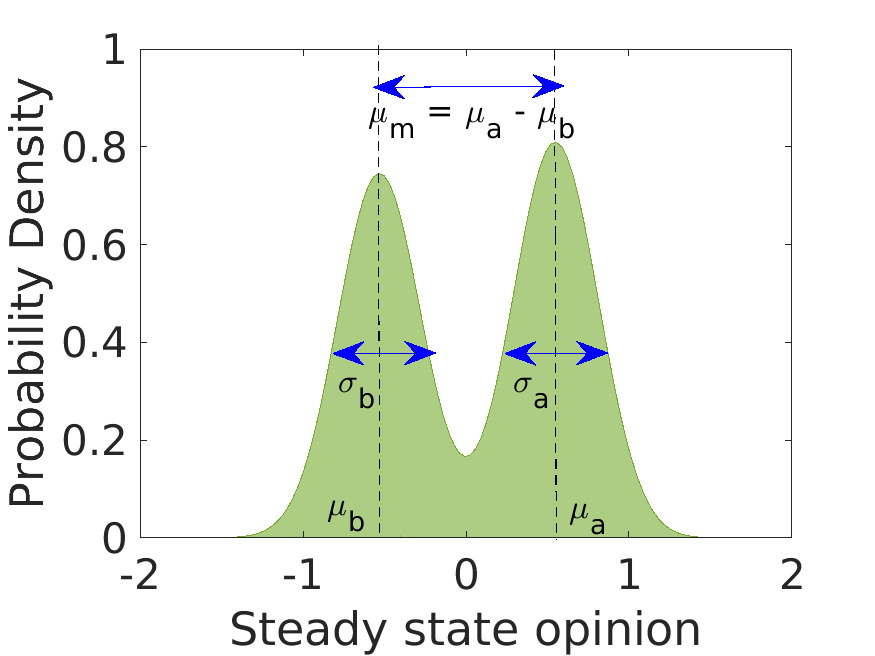}
    \includegraphics[width=0.32\textwidth]{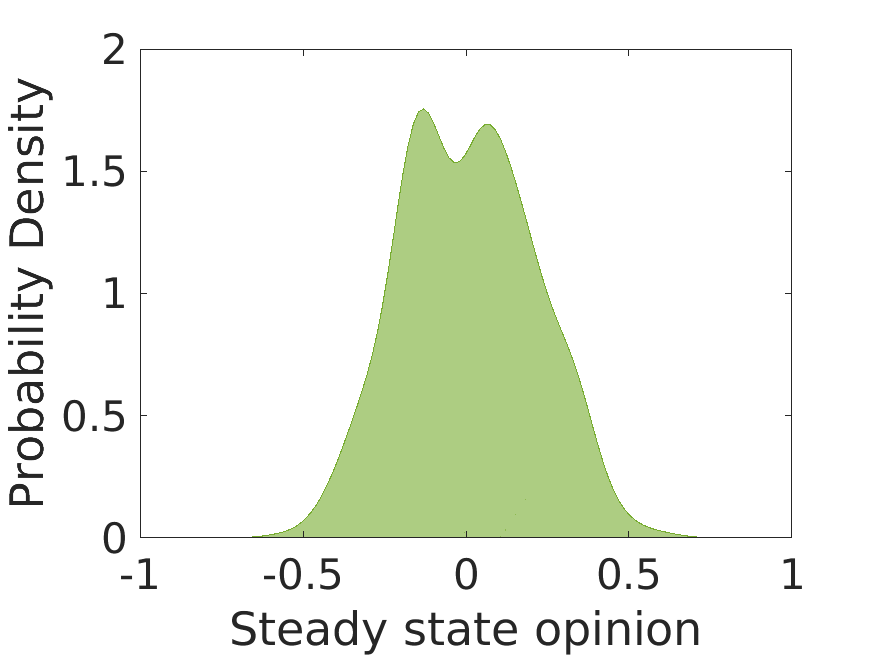}
    \includegraphics[width=0.32\textwidth]{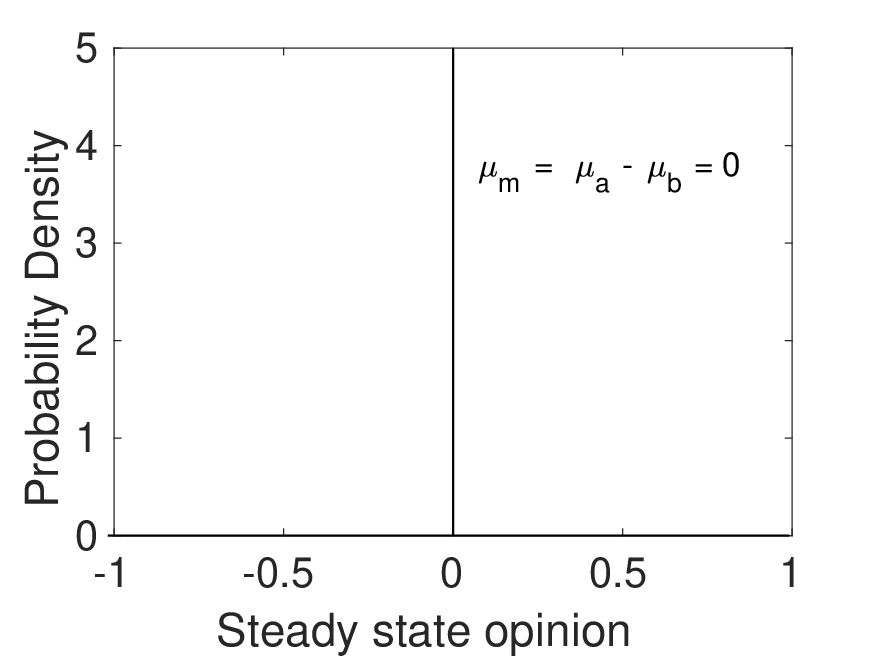}\\
    \textbf{(a)} \hspace{4.7cm} \textbf{(b)}\hspace{4.7cm} \textbf{(c)}\\
    \caption{Probability density function (pdf) of steady state values of opinions for (a) $F_m = -0.3$ (b) $F_m = -0.5$ (c) $F_m = -0.9$. Other parameters are $\alpha_m = 0.07$, $\alpha = 3$ and $\beta = 3$.}
    \label{fig:fig3}
\end{figure}

\begin{figure}[htbp]
\centering
    \includegraphics[width=0.8\textwidth]{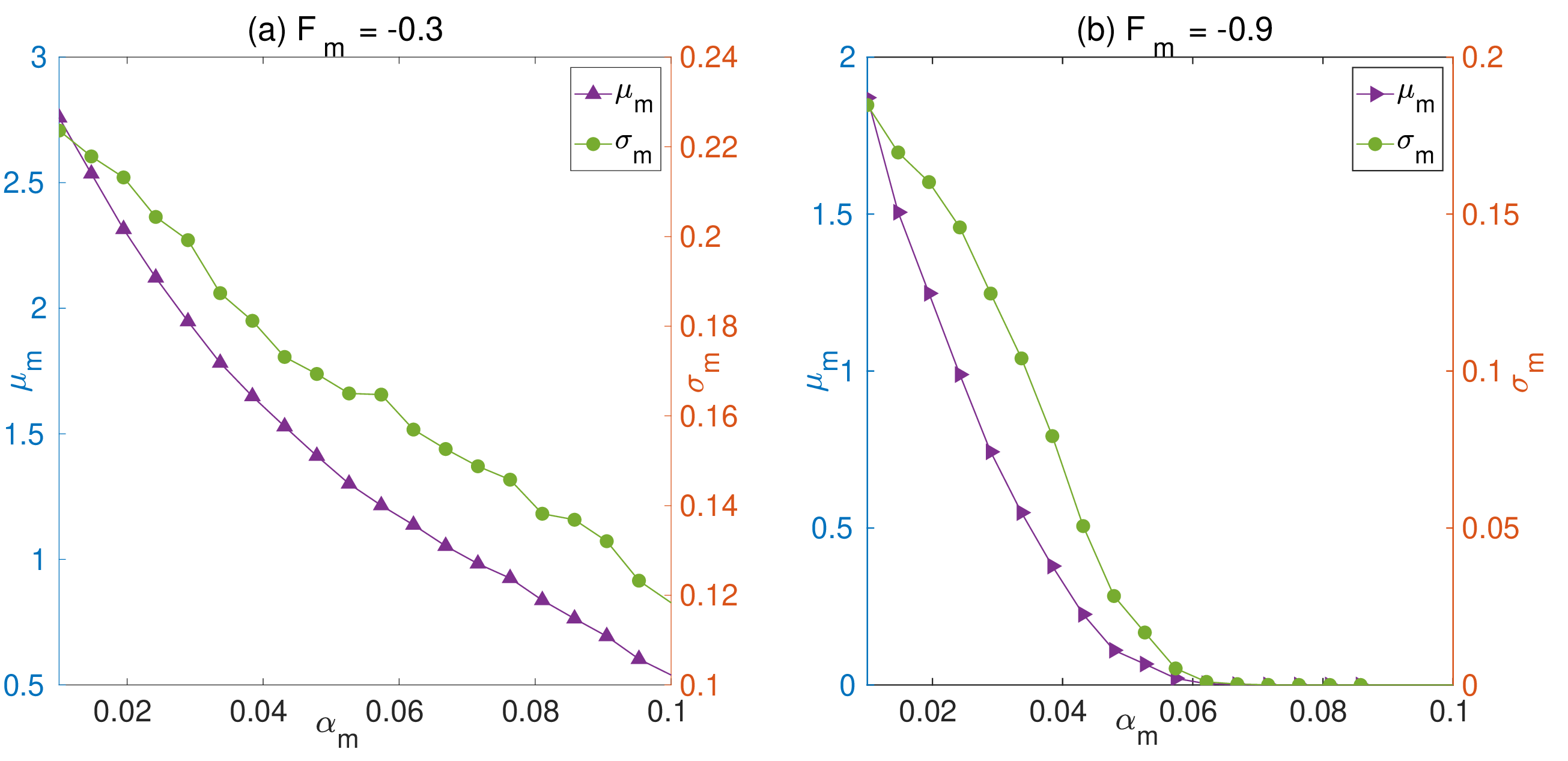}
    \caption{Variation of $\mu_m$ (left y-axis) and $\sigma_m$ (right y-axis) for a fixed value of $F_m$ with $\alpha_m$ for (a) $F_m = -0.3$ (b) $F_m = -0.9$.}
\label{fig:fig4}
\end{figure}

The collective dynamical behavior of the agents in the network can be analyzed  through the stationary probability density function (pdf) of the state variables, after sufficiently long time integration for the coupled equations of the network.  Fig. \ref{fig:fig3} shows the pdf for   a fixed value of $\alpha_m = 0.07$ and for different cases of $F_m$. From the figure it is observed that for less negative values of $F_m$, the resulting pdf is bimodal, indicating the formation of echo chambers in a polarized society; see Fig. \ref{fig:fig3}(a) . As $F_m$ becomes more negative, the two peaks in the pdf move towards each other (see Fig.\ref{fig:fig3}(b))  and finally results in a Dirac-Delta pdf as seen in Fig. \ref{fig:fig3}(c). The Dirac-delta pdf indicates the formation of consensus in the society. This transition in the pdf as $F_m$ is varied can be quantified by defining the following order parameters: Let $\mu_m = \mu_a - \mu_b$ denote the distance between the two peaks in the pdf \cite{currin2022depolarization}, where $\mu_a$ and $\mu_b$ are the position of the two peaks. $\sigma_a$ and $\sigma_b$ be the spread of opinions with respect to $\mu_a$ and $\mu_b$ respectively  and let $\sigma_m = \frac{\sigma_a + \sigma_b}{2}$. These order parameters are indicated in Fig. \ref{fig:fig3}(a). The order parameters $\mu_m$ and $\sigma_m$ for a fixed value of $F_m$ and varying $\alpha_m$ is shown in Fig. \ref{fig:fig4}.
%
%
%
%
%
As the value of $\alpha_m$ is increased, both $\mu_m$ and $\sigma_m$ decreases. This indicates that the two peaks in the bimodal pdf move closer to each other. Additionally, it can be observed that the spread of opinions about the mean value $\mu_a$ and $\mu_b$ decreases {for $F_m=-0.3$}; see Fig. \ref{fig:fig4}(a). {But for $F_m=-0.9$}, as the value of $\alpha_m$ is increased, $\mu_m$ and $\sigma_m$ decreases and finally makes a transition to zero as shown in Fig. \ref{fig:fig4}(b). This zero value indicates a Dirac-delta pdf as seen in Figure \ref{fig:fig3}(c). In summary, these order parameters, $\mu_m$ and $\sigma_m$, quantify the convergence or divergence of opinions within a society influenced by social media and individual openness to diverse perspectives. As individuals become more receptive to opposing views, societal segregation decreases, leading to a narrowing of opinion differences and, ultimately, to consensus.

\begin{figure}[t]
\centering
    \includegraphics[width=6.0in,height=4.2in]{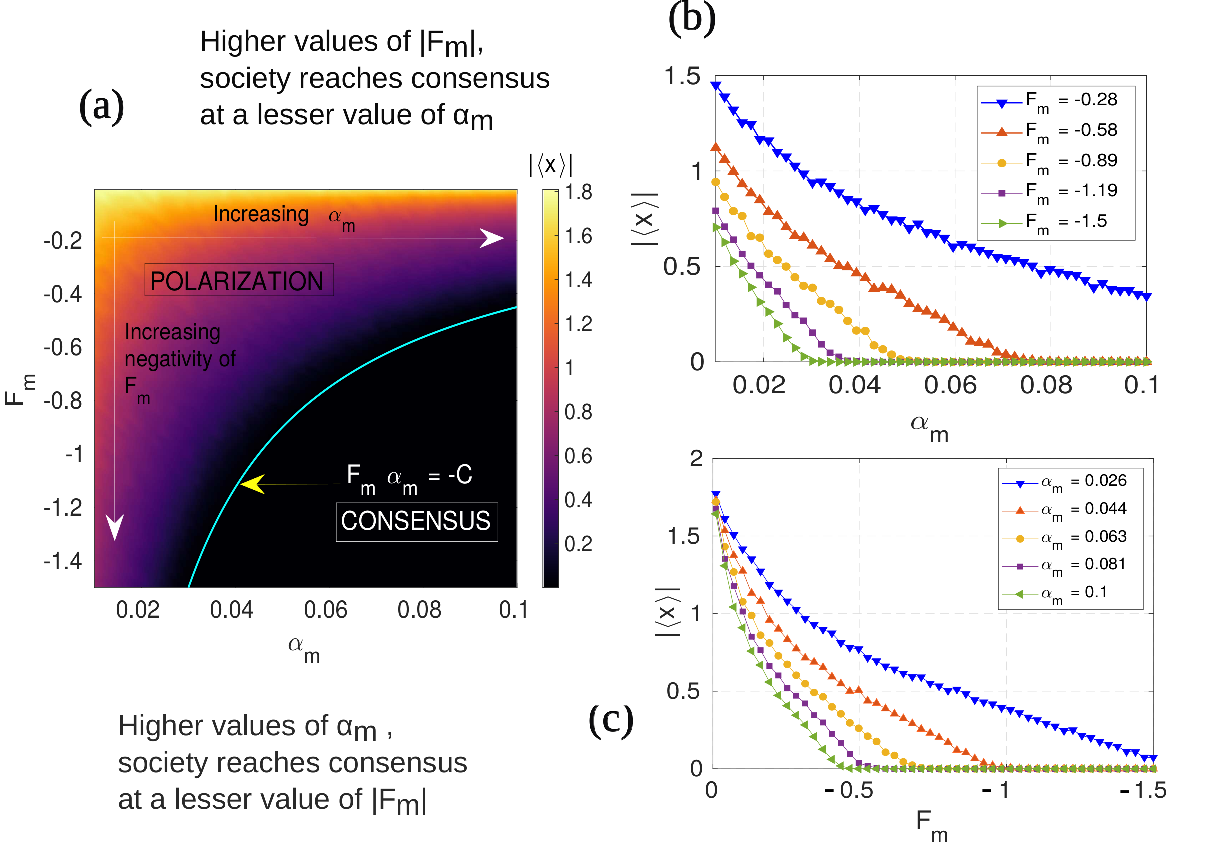}
    \caption{(a) Plot of $|\langle x \rangle|$ for different values of $(\alpha_m,F_m)$. The white line is the analytical expression $F_m\alpha_m = -C$ (refer Equation \ref{transition1}) obtained for the boundary which separates polarization from consensus. (b) Plot of $|\langle x \rangle|$ with $\alpha_m$ for different values of $F_m$. (c) Plot of $|\langle x \rangle|$ with $F_m$ for different values of $\alpha_m$.  }
\label{fig:fig5}
\end{figure}

The evolution of opinions for a wide range of values of $\alpha_m$ and $F_m$ in the parameter space is next explored. For each point in the parameter space in the  $\alpha_m-F_m$ plane, Eq.\eqref{media} is numerically integrated till the system reaches the steady state. At the steady state time $t$, average of the absolute opinions $|\langle x \rangle| = \frac{1}{N}\sum _{i=1}^{N}|x_i|_t$ is calculated. The metric $|\langle x \rangle|$ is treated as the order parameter for further investigations. It is observed that by increasing both $\alpha_m$ and $|F_m|$, the society undergoes a transition from polarization to consensus. A polarized state is obtained when $|\langle x \rangle| > 0$,  whereas $|\langle x \rangle| = 0$ implies a state of consensus. Fig. \ref{fig:fig5} showcases the results obtained from this analysis.
Fig.\ref{fig:fig5}(a) shows the parameter plot of $|\langle x \rangle|$ obtained for a wide range of values in the parameter space  $\alpha_m- F_m$. When $F_m \in [-0.4,-0.1]$, an increase in $\alpha_m$ leads to a gradual decrease in $|\langle x \rangle|$. However, when $F_m \in [-1.5, -0.4)$, further increments in $\alpha_m$ eventually reaches a threshold where $|\langle x \rangle|$ becomes zero, indicating a transition from polarized state to consensus. This result can also be observed in Fig. \ref{fig:fig5}(b), where $|\langle x \rangle|$ is plotted for different values of $F_m$ against $\alpha_m$. Note that for increasingly negative values of $F_m$, the threshold of $\alpha_m$ required for $|\langle x \rangle|$ to reach zero decreases.  Fig. \ref{fig:fig5}(c) specifically displays the relationship of $|\langle x \rangle|$ as $F_m$ varies across different $\alpha_m$ values. Notably, the value of $F_m$ for which the system transitions from polarization to consensus shifts to the left (see Fig. \ref{fig:fig5}(c)) when $\alpha_m$ is set at higher levels.


Next, the critical transition points indicating the shift from polarization to consensus triggered by the influence of social media effects is derived using the procedure outlined in 
\cite{baumann2020modeling}. 
To obtain the transition point, the Jacobian $\mathbf{J}$ of Eq.\eqref{final} is evaluated at the fixed point $\mathbf{x = x^* = 0}$, and is given by
\begin{align}
    \mathbf{J_{x^* = 0}} = 
    \begin{bmatrix}
    -1 & Kc\alpha - F_m\alpha_m & \dots  & Kc\alpha - F_m\alpha_m \\
    Kc\alpha - F_m\alpha_m & -1 & \dots  & Kc\alpha - F_m\alpha_m \\
    \vdots &  \vdots & \ddots & \vdots \\
    Kc\alpha - F_m\alpha_m & Kc\alpha - F_m\alpha_m  & \dots  & -1
\end{bmatrix}.
\end{align}
The largest eigenvalue of $\mathbf{J}$ denoted by $\zeta_{max}$ is given as
\begin{align}
    \zeta_{max} = -1 + (N-1)Kc\alpha - F_m\alpha_m = -1 + \frac{N-1}{N}K\alpha m(1+r)\langle a \rangle - F_m\alpha_m.
\end{align}
As the consensus state destabilizes when $\zeta_{max}$ becomes greater that zero, $\zeta_{max} = 0$ is the critical case. This leads to the following critical condition
\begin{align}
\label{transition}
   F_m\alpha_m = -1 + \frac{N-1}{N}K\alpha m(1+r)\langle a \rangle, 
\end{align}
and in the limit of $N \to \infty$, Eq. \eqref{transition} leads to the condition
\begin{align}\label{transition1}
   F_m\alpha_m = -1 + K\alpha m(1+r)\langle a \rangle = -(1-K\alpha m(1+r)\langle a \rangle)
\end{align}
The above equation can be written in a compact form as $F_m\alpha_m = -C$ where $C =  1 - K\alpha m(1+r)\langle a \rangle $. This equation which separates the state of network polarization from consensus is plotted in Figure \ref{fig:fig5}(a), shown as a cyan line, and validates the results obtained through numerical simulations. In the absence of social media effects ($F_m = 0$), Eq.(3) in \cite{baumann2020modeling} is obtained. 

\subsection{Positive media influence: $F_m > 0$}
\begin{figure}[htbp]
\centering
    \includegraphics[width=0.20\textwidth]{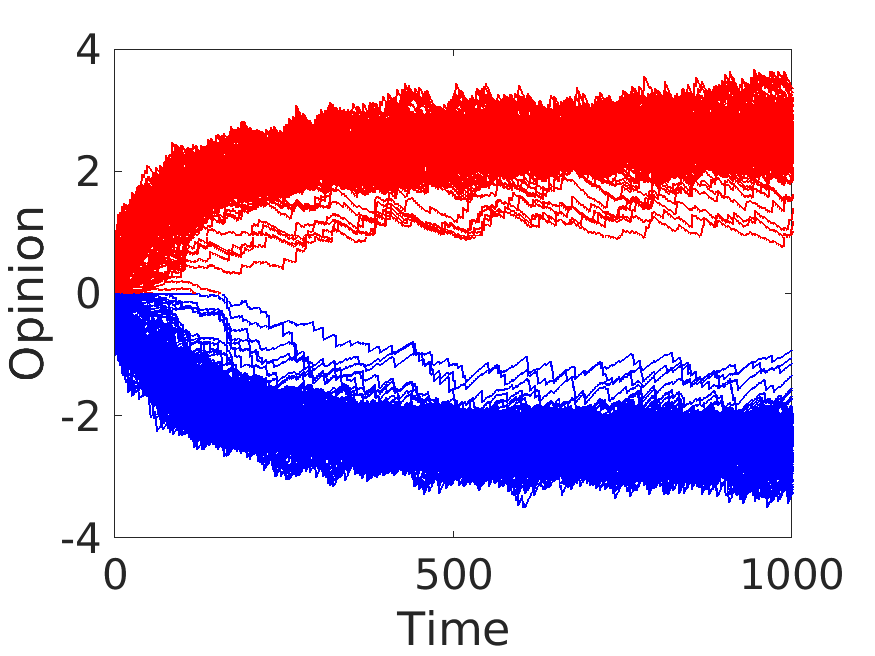}
    \includegraphics[width=0.20\textwidth]{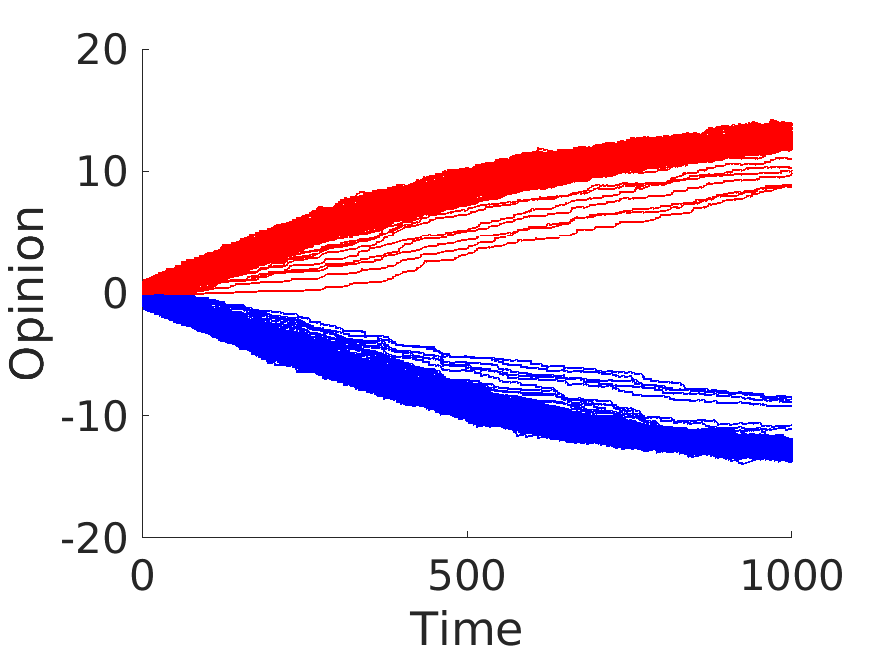}
    \includegraphics[width=0.20\textwidth]{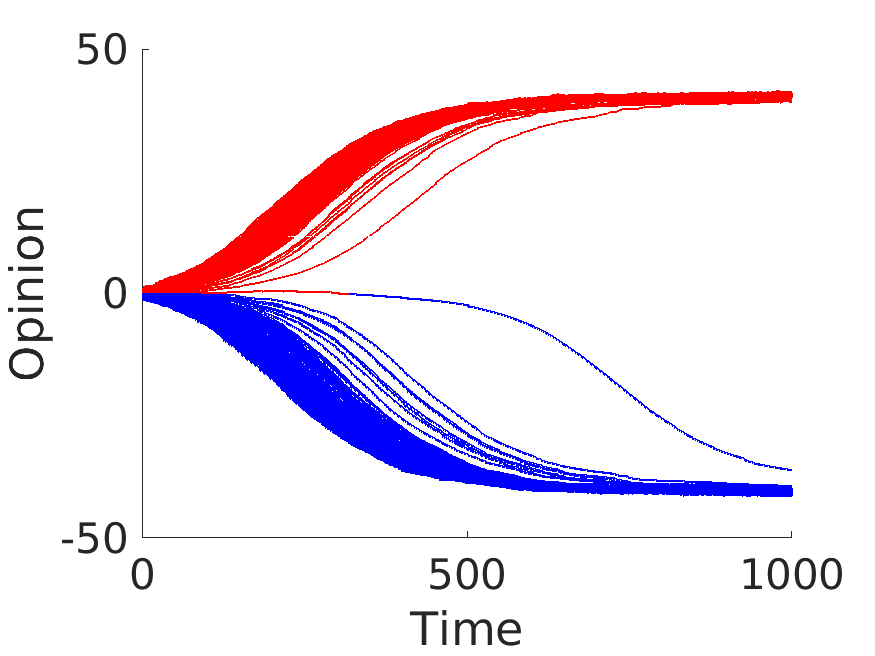}\\
    \textbf{(a)} \hspace{3.2cm} \textbf{(b)}\hspace{3.2cm} \textbf{(c)}\\
    \includegraphics[width=0.20\textwidth]{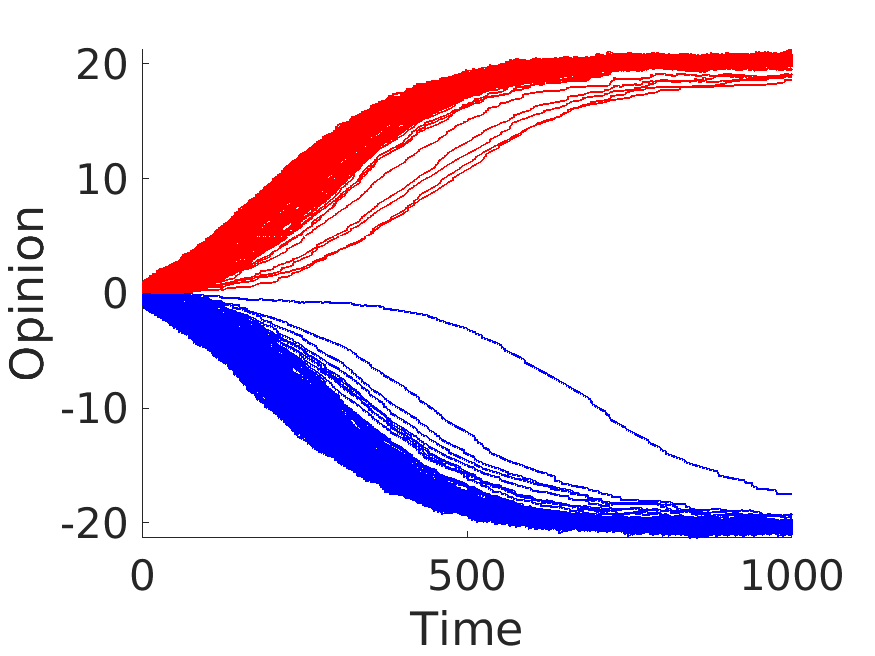}
    \includegraphics[width=0.20\textwidth]{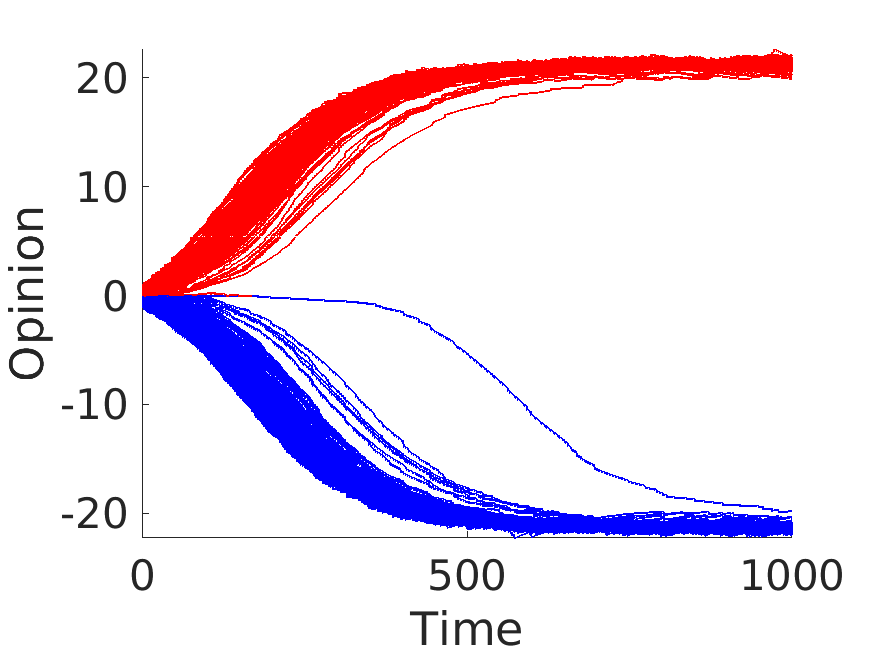}
    \includegraphics[width=0.20\textwidth]{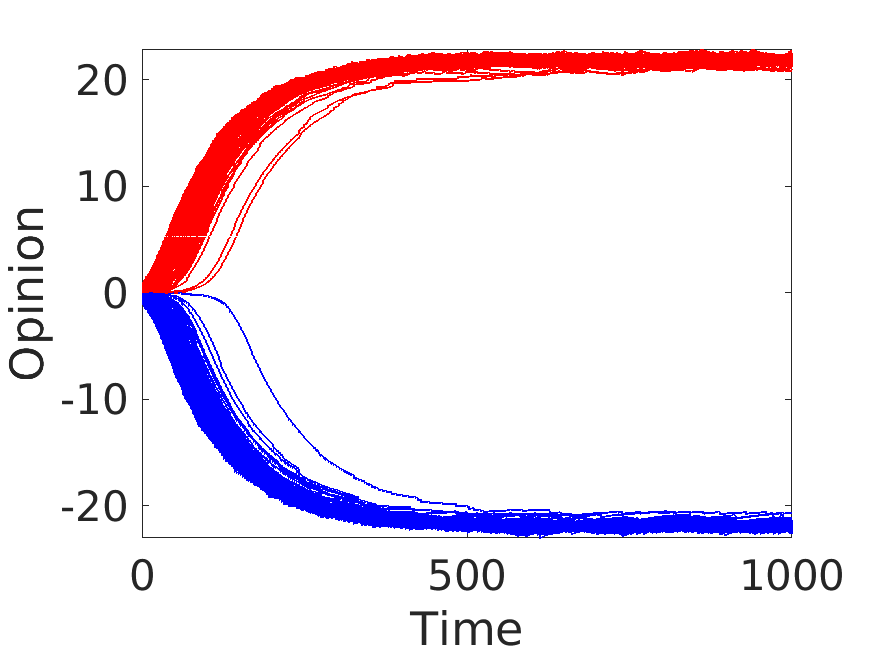}\\
    \textbf{(d)} \hspace{3.2cm} \textbf{(e)}\hspace{3.2cm} \textbf{(f)}\\
    \caption{Temporal evolution of the agents’ opinions for  (a) $F_m = 0.05$, $\alpha_m = 0.05$ (b) $F_m = 0.2$, $\alpha_m = 0.05$ (c) $F_m = 0.4$, $\alpha_m = 0.05$ (d) $F_m = 0.2$, $\alpha_m = 0.08$ (e) $F_m = 0.2$, $\alpha_m = 0.1$ (f) $F_m = 0.2$, $\alpha_m = 0.3$. Other parameters are $\alpha = 3$ and $\beta = 3$.}
    \label{fig:fig6}
\end{figure}

The case of $F_m>0$ is investigated next. Here, the media reinforces the existing opinions of the agents and therefore, the social media provides a reinforcing influence in a society that 
has the tendency to polarize. As in the earlier case, 
the numerical values of $\alpha$ and $\beta$ are taken to be $3$. 
Eq.\eqref{media} is numerically integrated for different values of $\alpha_m$ and $F_m$ and the results are shown in Fig. \ref{fig:fig6}. 
The top panel of Fig. \ref{fig:fig6} shows the evolution of  opinions of all the agents for three cases of $F_m$ for a fixed value of $\alpha_m = 0.05$. As can be seen, with an increase in  $F_m$, there is an observable rise in the steady-state opinions. Simultaneously, the divergence within the echo chambers diminishes, converging towards a scenario where opinions among all agents within each echo chamber tend to approach a common steady state opinion value. Different behaviour is observed when $\alpha_m$ is varied for a constant $F_m = 0.2$. From the bottom panel of Fig.\ref{fig:fig6}, it is observed that for higher values of $\alpha_m$, the steady state opinions remain same for all values of $\alpha_m$ but the time taken to reach the steady state reduces as $\alpha_m$ increases. 
\begin{figure}[htbp]
\centering
    \includegraphics[width=0.32\textwidth]{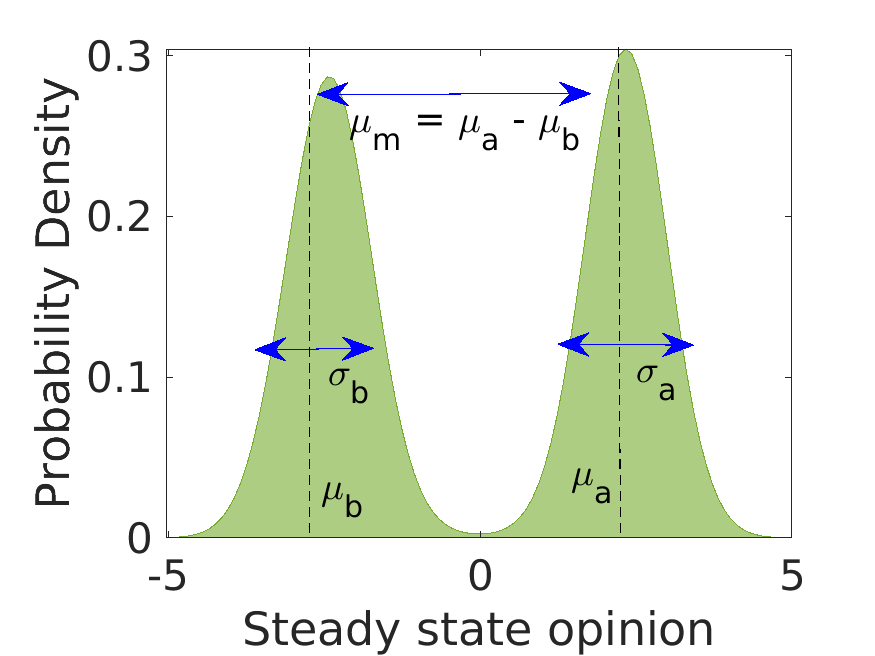}
    \includegraphics[width=0.32\textwidth]{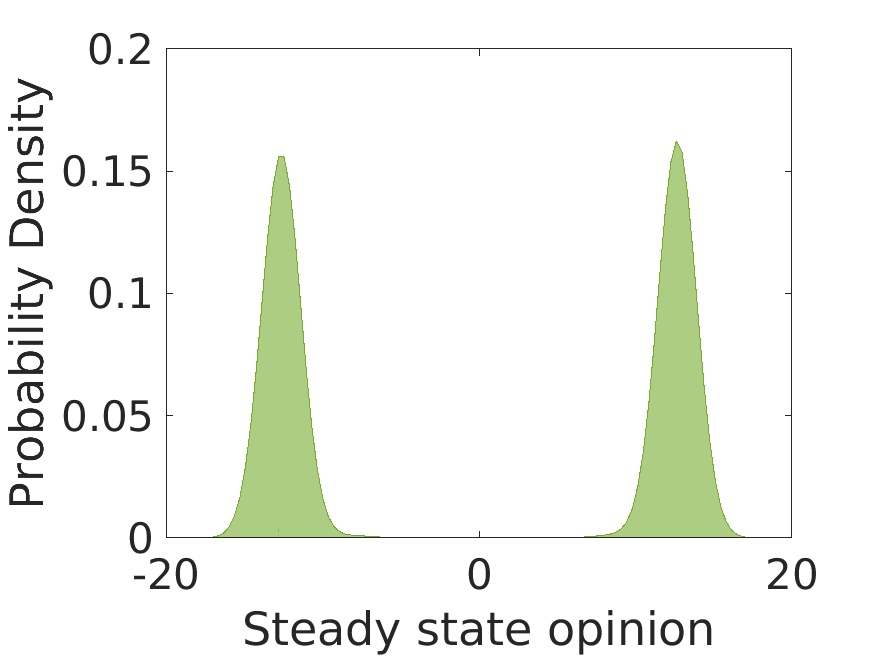}
    \includegraphics[width=0.32\textwidth]{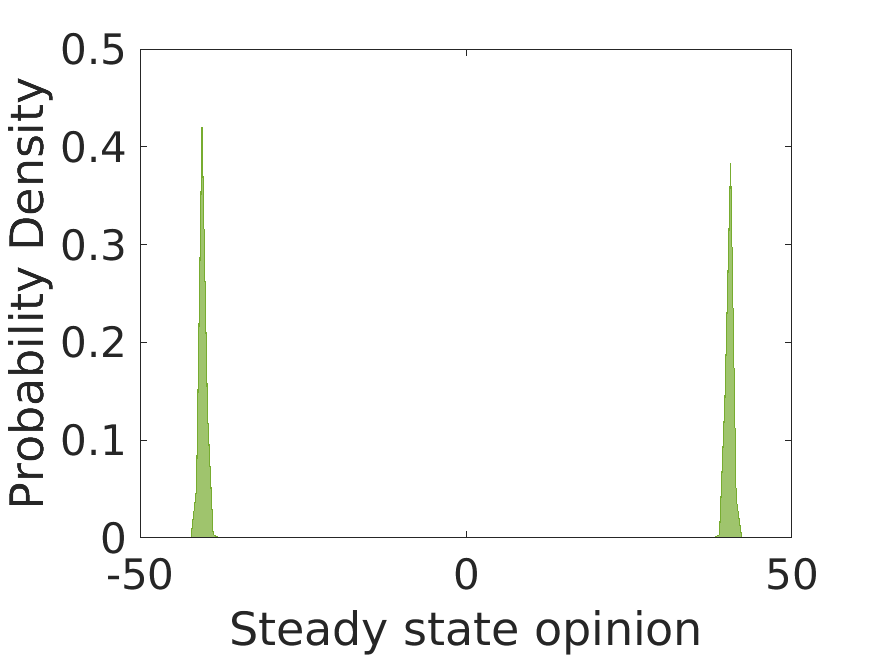}\\
    \textbf{(a)} \hspace{4.7cm} \textbf{(b)}\hspace{4.7cm} \textbf{(c)}\\
    \caption{Probability density function (pdf) of steady state values of $x_i$ for (a) $F_m = 0.05$ (b) $F_m = 0.2$ (c) $F_m = 0.4$. Other parameters are $\alpha_m = 0.05$, $\alpha = 3$ and $\beta = 3$.}
    \label{fig:fig7}
\end{figure}

The collective behaviour exhibited by varying $F_m$ and $\alpha_m$ can be better understood by plotting the pdf of the steady state values of the opinions. Fig. \ref{fig:fig7} shows the pdfs of steady state values of opinions plotted for different $F_m$ with $\alpha_m = 0.05$. Across all values of $F_m$, the pdf consistently shows a bimodal nature, signaling polarization within the society; this is in contrast to the behavior observed in Fig. \ref{fig:fig3}. Upon increasing the value of $F_m$, there is a noticeable amplification in the separation between the peaks, accompanied by an increased peak sharpness within the distribution. This is effectively captured by the parameters $\mu_a$, $\mu_b, \mu_m$, $\sigma_a$, $\sigma_b$ and $\sigma_m$ shown in Fig. \ref{fig:fig7}(a). 
\begin{figure}[htbp]
\centering
    \includegraphics[width=0.9\textwidth]{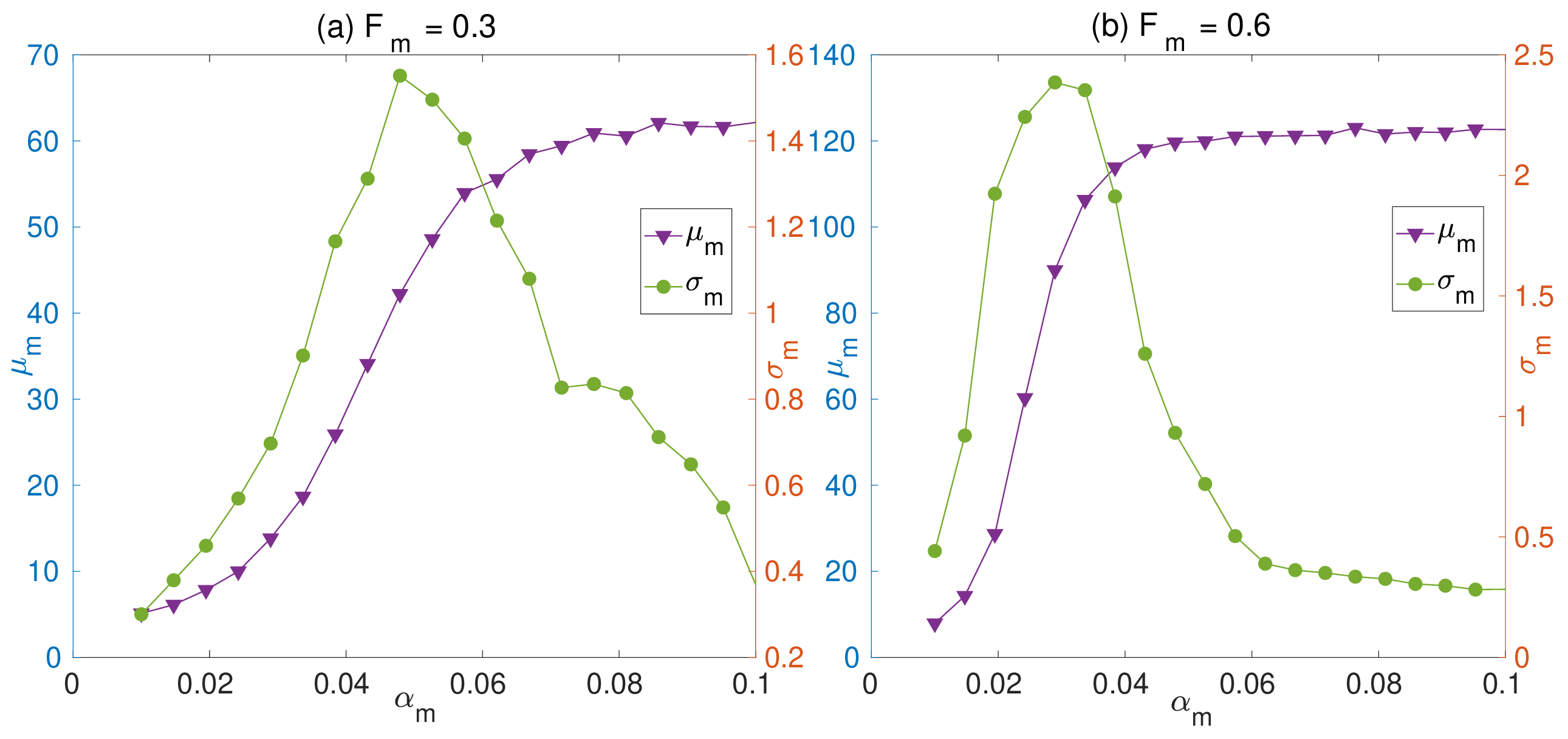}
    \caption{Variation of $\mu_m$ (left y-axis) and $\sigma_m$ (right y-axis) for a fixed value of $F_m$ with $\alpha_m$ for (a) $F_m = 0.3$ (b) $F_m = 0.6$.}
\label{fig:fig8}
\end{figure}
\begin{figure}[htbp]
\centering
    \includegraphics[width=0.9\textwidth]{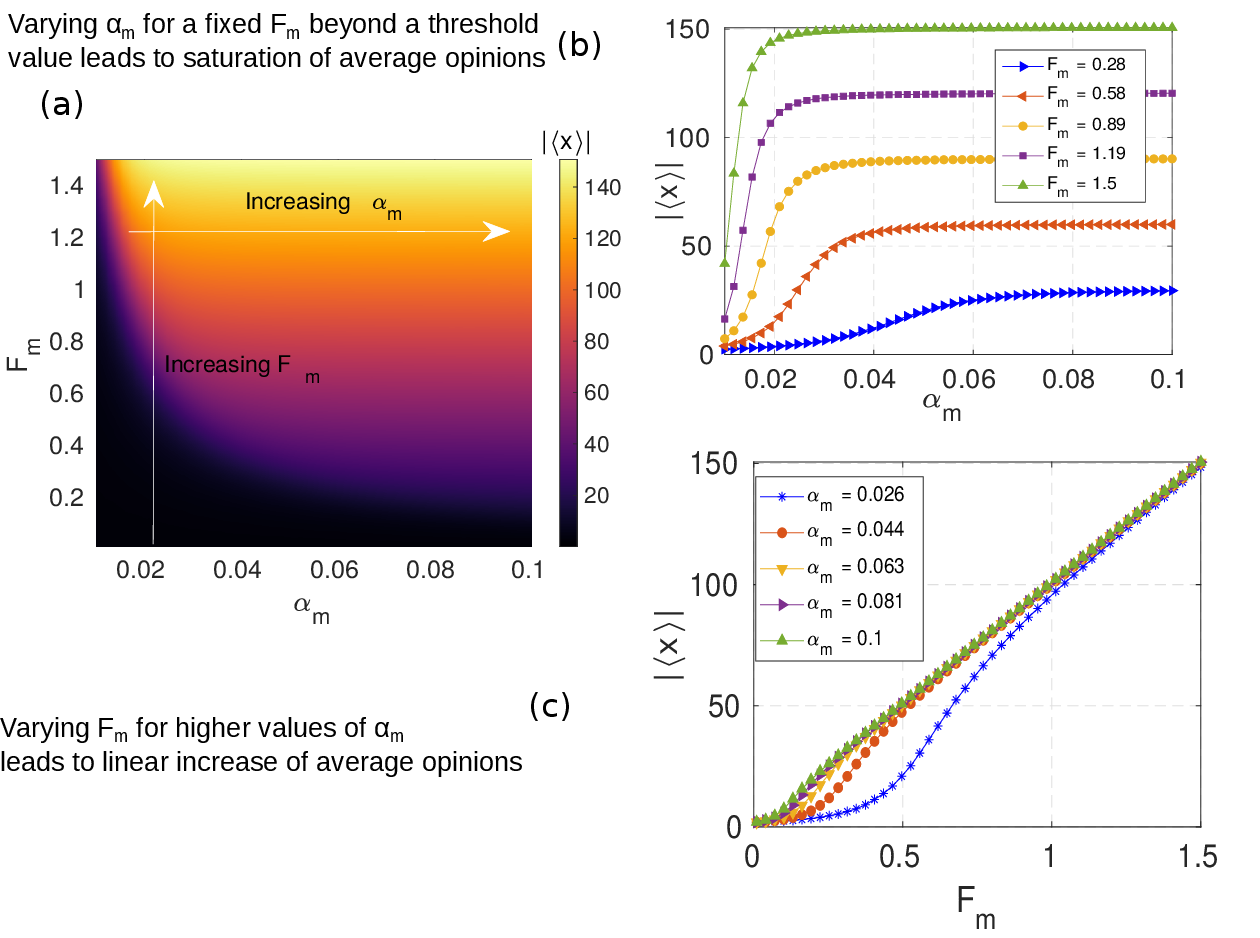}
    \caption{(a) Plot of $|\langle x \rangle|$ for different values of $(\alpha_m,F_m)$. (b) Plot of $|\langle x \rangle|$ with $\alpha_m$ for different values of $F_m$. (c) Plot of $|\langle x \rangle|$ with $F_m$ for different values of $\alpha_m$. }
\label{fig:fig9}
\end{figure}

Fig. \ref{fig:fig8} shows the variation of $\mu_m$ and $\sigma_m$ with $\alpha_m$ for two different values of $F_m$. As $\alpha_m$ increases, the parameter $\mu_m$, representing the distance between the peaks, initially increases before stabilizing at a constant value. However, a distinct trend is observed in $\sigma_m$. Initially, an increase in $\alpha_m$ results in an escalation of $\sigma_m$. This initial widening of $\sigma_m$ implies that as individuals become more receptive to varied viewpoints, the range of opinions within each echo chamber becomes broader. The consistent value of $\mu_m$ is counterbalanced by a decline in $\sigma_m$. This decrease means that although the distance between the peaks remains constant, the opinions within these clusters become more concentrated or tightly packed around each peak.

The findings presented in Fig. \ref{fig:fig6} are based on specific values of $\alpha_m$ and $F_m$. To gain a comprehensive understanding of how positive media influence ($F_m > 0$) impacts societal polarization, an analysis is conducted across the $\alpha_m$- $F_m$ parameter space. The range of values of $\alpha_m$ and $F_m$ is taken to be $[0.01,0.1]$ and $[0.01,1.5]$ respectively. For each point in the $\alpha_m-F_m$ parameter space,  Eq.\eqref{media} is evaluated till the steady state is reached. Once the system reaches steady state, $|\langle x\rangle|$ as defined in Fig. \ref{fig:fig5} is evaluated for every point in the $\alpha_m-F_m$ parameter space and the resulting plot is shown in Fig. \ref{fig:fig9}(a). The color code shown in Fig. \ref{fig:fig9}(a) represents the value of $|\langle x\rangle|$ calculated at steady state. For a constant $F_m$ and varying $\alpha_m$, the behavior of $|\langle x\rangle|$ shows a progressive increase until it reaches a particular $\alpha_m$ value, upon which it stabilizes. This trend is depicted in Fig. \ref{fig:fig9}(a), where, initially, there is a shift in color gradation for varying $\alpha_m$, followed by a stable color representation; refer to the colorbar for $|\langle x\rangle|$ values. A higher $F_m$ corresponds to a greater saturated value for $|\langle x\rangle|$. These observations are also presented in Fig. \ref{fig:fig9}(b), where $|\langle x\rangle|$ is plotted for various values of $F_m$ with varying $\alpha_m$. It should also be noted that, the threshold value of $\alpha_m$ at which $|\langle x\rangle|$ stabilizes is less for a higher value of $F_m$. 

Next, the change in $|\langle x\rangle|$ for a fixed $\alpha_m$ and varying $F_m$ is analysed. As shown in Fig. \ref{fig:fig9}(a), the progressive rise in $F_m$ corresponds to a consistent increase in $|\langle x\rangle|$. This trend is visually apparent through the transition in color from black to yellow ( from bottom to top) , indicating a shift from lower to higher values. In Fig. \ref{fig:fig9}(c), $|\langle x\rangle|$ is plotted for different values of $\alpha_m$ against $F_m$. For higher values of $\alpha_m$, increase in $|\langle x\rangle|$ with $F_m$ is approximately linear (see green line in Fig. \ref{fig:fig9}(c)) whereas, for lower values of $\alpha_m$, the linear trend is observed for higher values of $F_m$. This suggests that individuals who are more open-minded tend to be receptive to media input and are influenced even with relatively weaker media impact. Conversely, individuals less open to new perspectives  require a stronger media influence for alignment with the  viewpoint in the social media.

\section{Conclusion} \label{conclusion}
 The interaction patterns on social media platforms often foster polarization and the formation of echo chambers. For instance, research indicates that recommendation algorithms used by popular social media platforms like YouTube can intensify political polarization. Furthermore, studies suggest that when individuals have significant control over the information they consume, it tends to deepen partisan divisions within society \cite{stroud2010polarization}. This phenomenon arises from people's inclination to seek information that aligns with their existing beliefs. Given the significant influence of social media in shaping public opinion, this study uses a recent opinion dynamics model that incorporates the concept of homophily. The aim is to investigate how social media usage affects a society prone to polarization. The findings reveal that when individuals select information based solely on their preferences, it reinforces existing polarization. However, when individuals remain receptive to ideas which are contradictory to their own beliefs, a shift toward consensus is observed via supercritical pitchfork bifurcation. This underscores the dual nature of social media, which has the potential to either mitigate or amplify polarization within society. Consequently, individuals consuming information on social media should exercise caution and not solely rely on recommendation algorithms. Additionally, this study reveals that a society that is less zealot and more open to ideas that are divergent from their own beliefs are less likely to lead to a polarized society.

\section*{Acknowledgments}
The authors would like to acknowledge the funding received from the Ministry of Education, Government of India towards Center on Complex Systems and Dynamics under the Institute of Eminence scheme.

\section*{Author Declarations}
\subsection*{Conflict of Interest}
The authors have no conflicts of interest to declare.

\subsection*{Author Contributions}
{\bf SP}:Conceptualization (equal);  Formal analysis (equal); Investigation (lead); Methodology (equal);  Software (lead); Validation (equal); Visualization (lead); Writing – original draft (equal).

{\bf SG}: Conceptualization (equal); Funding acquisition (lead); Investigation (supporting); Methodology (supporting); Project administration (lead); Supervision (lead); Visualization (supporting); Writing – review \& editing (lead).

\section*{Data Availability} No data has been used in this study.

\bibliographystyle{ieeetr}
\bibliography{references}

\end{document}